\documentclass{article}

\usepackage[
	hidelinks,
	colorlinks = true,
	linkcolor  = {blue!50!black},
	citecolor  = {blue!50!black}
]{hyperref}
\usepackage{cite}
\usepackage{comment}
\usepackage{enumitem}
\usepackage{amsmath}
\usepackage{amssymb}
\usepackage{siunitx}
\usepackage{graphicx}
\usepackage{tikz}
\usepackage{tikz-3dplot}
\usepackage{pgfplots}
\pgfplotsset{compat=1.14}
\usepackage{nicefrac}
\usepackage{tabularx}
\newcolumntype{C}{>{\centering\arraybackslash}X}
\usetikzlibrary{calc}
\usetikzlibrary{fadings}
\usetikzlibrary{arrows.meta} 

\tikzstyle{textoverlaywhite} = [
	inner sep=1pt, align=left,
	text=white, font=\footnotesize,
	fill=black, rounded corners=2pt,
	fill opacity=0.40, text opacity=1
]

\newcommand\drawlight[1]{
	\fill [black!20!yellow] (#1) circle [radius=0.6pt];
	\foreach \x in {0,20,...,350}
		\draw [thin, black!20!yellow] ($(#1) + (\x:0.04)$) -- +(\x:0.1);
}

\newcommand\draweyex[1]{
	\def\eyeborder{($(#1)+(-7pt,-1pt)$) to[out=50,in=170] +(7pt,3pt) to[out=290,in=60] +(-0.4pt,-4pt) to[out=180,in=350] +(-6.6pt,1pt) };
	
	\begin{scope}
		\clip\eyeborder;
		\fill [green!40!blue] (#1) circle [radius=2.8pt];
		\foreach \a in {1,24,76,154,199,260,302}
			\draw [green!57!blue] ($(#1) + (\a:0.7pt)$) -- +(\a:2pt);
		\foreach \a in {53,111,177,219,341}
			\draw [green!45!blue] ($(#1) + (\a:0.7pt)$) -- +(\a:2pt);
		\fill [black] (#1) circle [radius=0.8pt];
	\end{scope}
	
	\draw[black!75,-{Triangle Cap[length=0.5pt]}] ($(#1)+(-7pt,-1pt)$) to[out=50,in=170] +(7.2pt,2.95pt) to[out=350,in=190] +(1.1pt,-0.05pt);
	\draw[black!75, line cap=round] ($(#1)+(-0.3pt,-2pt)$) to[out=180,in=350] +(-6.7pt,1pt);
}

\pgfplotsset{convergenceplot/.style = {
	width = 9.6cm,
	height = 6.5cm,
	name = spheres,
	anchor = south west,
	y label style = {at={(axis description cs:0.125,.5)},anchor=south},
	y tick label style = {rotate=90, anchor=south},
	xlabel style = {at={(axis description cs:0.5,-0.02)},anchor=north},
	legend cell align = left,
	legend style = {font=\scriptsize, at={(0.94, 0.93)}},
	xlabel = {iterations $N$},
}}

\tikzset{
	photon/.style = {inner color=yellow,outer color=CLight,draw=CLight,ultra thin},
	nee/.style = {inner color=orange,outer color=red,draw=red,ultra thin},
	>=latex
}

\newcommand\mvec[1]{\mathbf{#1}}

\usepackage{xcolor}
\colorlet{CLight}{black!20!yellow}

\makeatletter
\let\@orcid\relax
\def\orcid@base{https://orcid.org/}
\def\orcid#1{\hbox{\href{\orcid@base #1}{\includegraphics{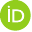}}}}
\makeatother

\usepackage[procnames]{listings}
\definecolor{lstcomment}{rgb}{0.0, 0.4, 0.03}
\lstdefinelanguage{pseudo}{
    frame        = lines,
    aboveskip    = \bigskipamount,
    belowskip    = \bigskipamount,
	sensitive	 = true,
	commentstyle = \color{lstcomment}\textit,
	morecomment	 = [l]{//},
	morecomment	 = [n]{/*}{*/},
	tabsize		 = 2,
	keywordstyle = \color{blue},
	stringstyle	 = \color[gray]{0.4},
	basicstyle	 = \fontsize{7.6}{7.6}\fontfamily{pcr}\selectfont,
	morekeywords = {for, if, while, switch, case, in, end, else, elif, cases, do, return, vec3, out, ref, match, uint16, uint32, uint64, int, ivec3, break, true, false, func, u32, f32, i32, class, inline, float, bool, void, template, typename, static, const, constexpr, nullptr, private, public},
	morestring   = [b]",
	mathescape   = true,
	keywordstyle = [2]{\color{orange!40!black}}, 
	morekeywords = [2]{sqrt,cos,sin,normalize,log,min,max, ceil, abs, dot, sign, atomic_add, compare_exchange_weak, swap, load, store, atomic_max, intersection_area, set_iteration, initialize, clear_counters, increment, get_density, get_density_robust, increment_if_positive, split_node_if_necessary, capacity, size},
	procnamekeys = {func}, 
	procnamestyle= \color{orange!50!black},
}
\title{Next Event Backtracking}
\author{Jendersie, Johannes \orcid{0000-0003-0703-4433}\\{\normalsize TU Clausthal, Germany}}

\begin{document}

\maketitle

\begin{figure}[h]
  \includegraphics[width=0.497\linewidth]{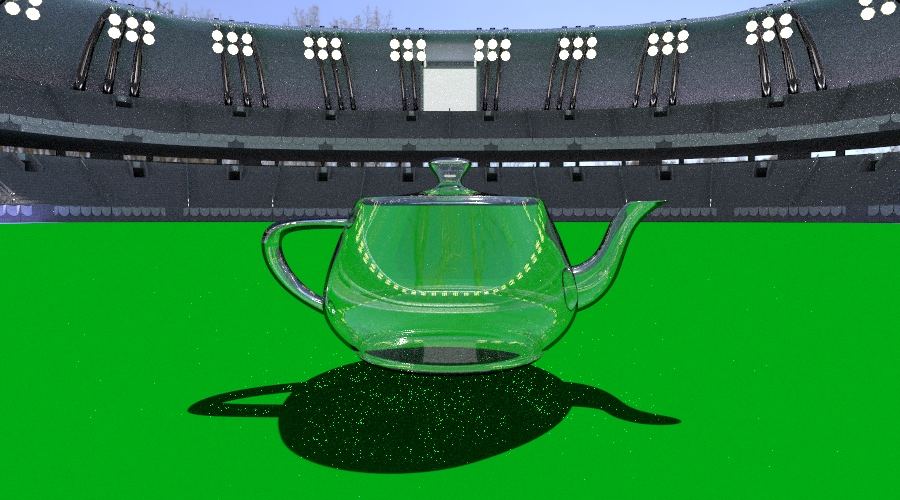}~%
  \includegraphics[width=0.497\linewidth]{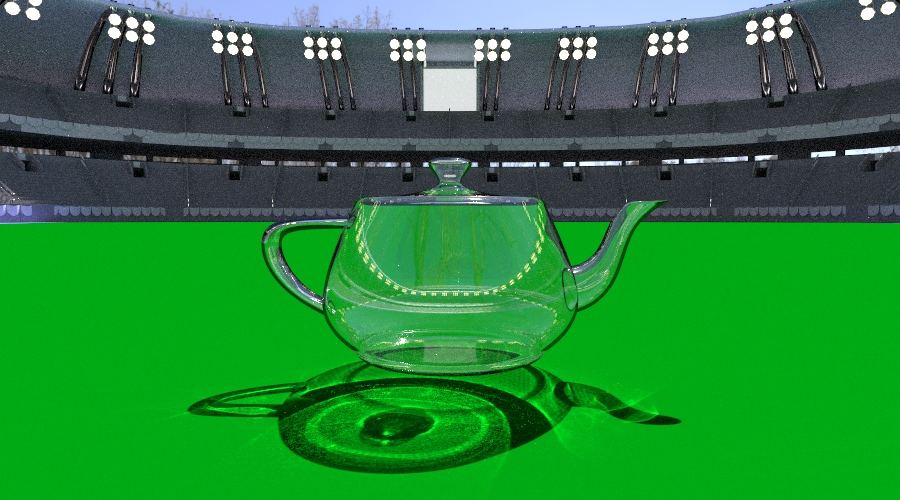}
  \centering
  \caption{Bidirectional Path Tracing (left, 212 spp) compared to Path Tracing with Next Event Backtracking (341 spp, equal time 15 \si{min}).
  Bidirectional Path Tracing and many more methods fail to produce the caustic in this scenario, because the light paths rarely hit the small teapot.
  Our method implicitly guides the light paths to the important regions and scales well for large scenes and many lights.
  }
  \label{fig:teaser}
\end{figure}

\begin{abstract}
  In light transport simulation, challenging situations are caused by the variety of materials and the relative length of path segments.
  Path Tracing can handle many situations and scales well to parallel hardware.
  However, it is not able to produce paths which have a smooth surface in connection with a small light source.
  Here, photon transports perform superior, which can be ineffective if the smooth object is small compared to the scene size.
  
  We propose to use the last segment of a Path Tracer path as the first segment of a photon path.
  As a result, the strengths of next event estimation are inherited by the photon transport and photons are guided toward the regions where they are most useful.
  To that end, we developed a lock-free sparse octree, which we use for fast and robust density estimates.
  Summarizing, the new method can outperform state of the art algorithms like Vertex Connection and Merging in certain scenarios.
\end{abstract}  


\section{Introduction}
\label{sec:introduction}

In stochastic light transport simulation there are three basic operators to create light paths: \textit{random walks}, \textit{connections} and \textit{merges}.
Random walks start at the observer or a light source and use ray tracing to find the next intersection point (vertex) in a randomly sampled direction.
If they hit a surface with the adjoint quantity, a full light transport path is found.
Random walks have a high chance for finding contributions if there are large light sources and/or the path to the light source is close to deterministic
(for example, if the emitter is directly visible or reflected by mirrors).

Connections can be created from each vertex of a random walk path.
By choosing a random point on an adjoint surface,
a contribution can be computed over arbitrary large distances.
This is called \textit{Next Event Estimation} (NEE), because the next event in the random walk could be the very same location.
The strength of NEE is that it can find light sources with a very small solid angle, which have a much smaller probability to be found via random walk alone.
Additionally, it is possible to use binary trees to select connection points proportional to their expected contribution to reduce the variance.

Finally, merges combine the vertices of two random walks by assuming they hit the same point, although they might be separated by a small distance.
That means, if the end points of two paths have a distance smaller than some threshold, they are combined to a full contribution path.
In general, merges have one more random sampling event than random walks and connections, leading to a higher variance in the first place.
However, their strength comes from the fact that neighbor points can be searched in the large set of vertices over all paths (often millions).
Merges are the only really successful operator to find the difficult specular-diffuse-specular (SDS) paths.

We combine connections and merges by using the connection segment as the first path segment of a new light walk.
Thus, merges inherit the strengths of next event estimation.
Therefore, the vertices, from which the connections were initiated, are transformed into photon-emitting virtual light sources.
This requires an estimate of the density of such NEE vertices.
For that purpose, we introduce a new octree-based data structure which is faster than kd-tree-based neighborhood searches.
Independent of the used data structure, every density estimate over a finite area will be biased and noisy, leading to a small bias in our new method.

\section{Related Work}
\label{sec:related}

The foundation of our method is the \textit{Path Tracing} (PT) algorithm which goes back to Kayjia \cite{kajiya_rendering_1986}.
Today, it is widely used in production renderers \cite{fascione_path_2017} due to its simplicity and extensibility.
It combines the two operators random walk and NEE by weighting the two independent estimates with respect to their effectiveness for each individual path.
This combination of two or more samplers for the same quantity is called \textit{Multiple Importance Sampling} (MIS) in Monte Carlo-based simulations.
In graphics, the balance or power heuristic are used to compute the weights.
They were introduced by Veach and Guibas \cite{veach_optimally_1995} and extended to a large set of operators since then.

PT can handle many situations, but fails for caustics and SDS paths.
\textit{Bidirectional Path Tracing} (BPT) \cite{veach_bidirectional_1995,lafortune_bi-directional_1993} is a stronger method.
It uses a random walk from the observer and another one from the light source, and computes all connections between the vertices of the two paths.
It is able to find caustics, but still fails for SDS paths.

The only successful methods to efficiently find SDS paths are based on photon transports.
\textit{Photon mapping} \cite{jensen_global_1996} is able to produce caustic and SDS paths, but has problems with the scene scale.
Large scenes or small specular objects as well as many light sources can lead to highly noisy results.
Jensen already used a dedicated \textit{Caustic Map} where photons are explicitly send into directions of specular objects.
This will still fail if there are many large specular objects and requires a generalization for glossy objects.
Contrarily, NEB shoots photons into important regions independent of the material.

Similarly, the first importance based method to distribute photons was introduced by Peter and Pietrek \cite{peter_importance_1998}.
It uses piecewise constant functions, created from the importance information of a view path tracing pass, to distribute photons.
Due to this piecewise approximation the method has problems with highly glossy surfaces.
The major difference to NEB is that the guidance in NEB will happen implicitly.

A different approach to improve the photon map quality is to stir the deposition of photons.
Suykens and Willems \cite{suykens_density_2000} fixed the photon mapping density to a constant, which reduces the number of photons in bright regions.
Likewise, Keller and Wald \cite{keller_efficient_2000} used an importance map to control the density of stored photons.
However, both methods are still sampling a large number of photons paths and only reduce the number of stored photons, wasting the others.

Georgiev et al. \cite{georgiev_light_2012} and Hachisuka et al. \cite{hachisuka_path_2012} simultaneously introduced the MIS-weight computation to successfully combine BPT with merges.
The \textit{Vertex Connection and Merging} (VCM) algorithm is one of the most robust methods so far.
Still, it may fail for selected situations, where the MIS-weight underestimates the variance of certain merge events.
This issue was overcome by Jendersie et al. \cite{jendersie_improved_2018,jendersie_variance_2019} with a small change to the heuristic.
While we only demonstrate our novel operator -- the next event backtracking (NEB) -- in Path Tracing, it is possible to integrate it into BPT or VCM, too.
However, NEB is already capable of handling many complex light situations without doing so.

Another important variant of the random walk operator is the \textit{Markov Cain Monte Carlo} (MCMC) sampling.
Instead of generating independent random sampling events, MCMC samplers apply small mutations to the paths.
Then, the new mutations are randomly discarded or accepted with respect to a target function.
This allows MCMC samplers to generate distributions of an unknown function to reduce noise opposed to the na{\"i}ve walk.
For an overview of MCMC methods we refer to the survey of \v{S}ik et al. \cite{sik_survey_2018}.
It is thinkable to use NEB in connection with MCMC random walks, which we leave as future work.

\textit{Manifold Next Event Estimation} (MNEE) from Hanika et al. \cite{hanika_manifold_2015} is the most similar method to ours.
There, random connections, which are blocked by refractive surfaces, are iteratively moved on the surface until they form a valid path.
This iteration requires multiple expensive evaluations and shadow tests and is biased in case there is an ambiguity of multiple possible lights paths.
Finally, MNEE is not able to find caustics from mirrors and prisms, where the reflection surface is not blocking the direct connection between the caustic and the light emitter.
NEB shares the weakness in that scenario, but at least handles it to a better degree than MNEE.
We believe that NEB is more robust than MNEE in general.

Other methods which target the problem of small caustic throwing objects are based on guidance.
A general guidance method like the method from M{\"u}ller et al. \cite{muller_practical_2017} reduces the overall variance in all sampling events by steering the samplers into the direction of the adjoint quantity.
A method which explicitly targets the exploration of visible caustics was invented by Hachisuka and Jensen \cite{hachisuka_robust_2011}.
It uses MCMC sampling for the light paths with the visibility of caustics as target function.
Recently, Grittman et al. \cite{grittmann_efficient_2018} improved the caustic handling in large scenes by restricting the use of photons adaptively and learning a guidance information at the light emitter.
Other than these methods our NEB does not learn those connections over time.
Instead, it samples these cases explicitly with a much higher density.

\section{Next Event Backtracking}
\label{sec:neb}

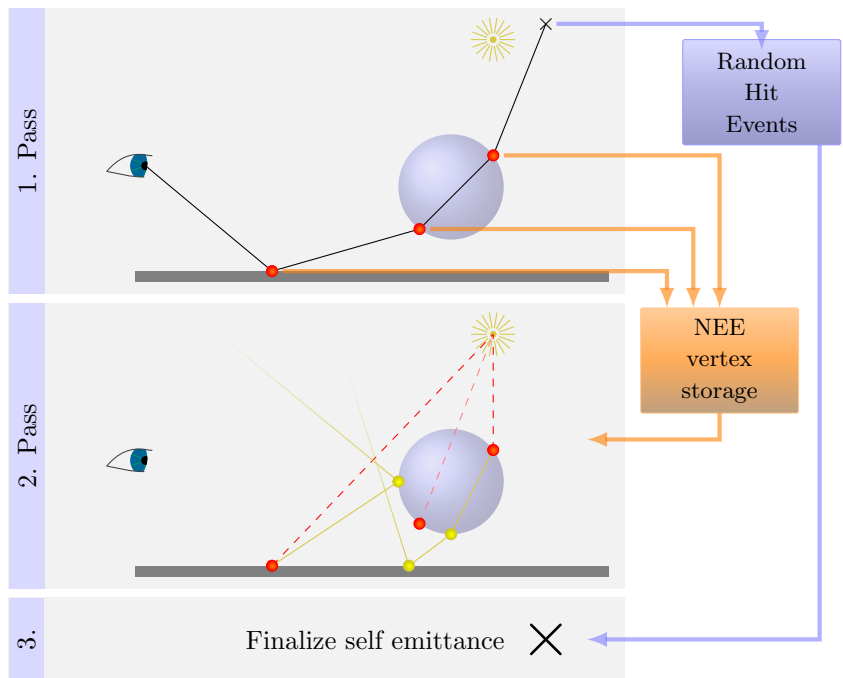
\begin{figure} \centering
\begin{tikzpicture}[scale=1.4]
  \def\scene{
	  \begin{scope}[xshift=1.3cm,yshift=-1.5cm,scale=1.5]
  		\draweyex{0,0};
 		\end{scope}
	  \begin{scope}[xshift=4.6cm,yshift=-0.3cm,scale=1.5]
  		\drawlight{0,0};
  	\end{scope}
  	\fill[black!50] (1.2,-2.6) rectangle +(4.5,0.1);
  	\fill[blue!30,shading=ball,ball color=blue!30, opacity=0.3] (4.2,-1.7) circle [radius=0.5];
  }
  
	\fill[black!5] (0,0) rectangle (5.85,-2.715);
	\node[fill=blue!15,rotate=90,anchor=north east,text width=3.56cm,align=center] at (0,0) {1. Pass};
	\scene
	\coordinate (p1) at (2.5,-2.5);
	\coordinate (p2) at (3.9,-2.1);
	\coordinate (p3) at (4.6,-1.4);
	\coordinate (p4) at (5.1,-0.15);
	\draw (1.3,-1.5) -- (p1) -- (p2) -- (p3) -- (p4);
	\node at (p4) {\small$\times$};
	\fill[nee] (p1) circle [radius=1.5pt];
	\fill[nee] (p2) circle [radius=1.5pt];
	\fill[nee] (p3) circle [radius=1.5pt];
	
	\draw[rounded corners=0.03cm, shade, top color=orange!40, bottom color=orange!50!black!50, middle color=orange!65, draw=orange!40] (6,-2.85) rectangle +(1.5,-1);
	\node[rotate=0,text width=1.5cm, align=center] at (6.75,-3.35) {\small NEE vertex storage};
	\draw[rounded corners=0.03cm, shade, top color=blue!15, bottom color=blue!50!black!40, middle color=blue!65!black!30, draw=blue!40] (6.4,-0.3) rectangle +(1.5,-1);
	\node[text width=1.5cm, align=center] at (7.15,-0.8) {\small Random Hit Events};
	
	\draw[->,orange,opacity=0.6,ultra thick] (2.6,-2.5) -- (6.25,-2.5) -- (6.25,-2.85);
	\draw[->,orange,opacity=0.6,ultra thick] (p2)+(0.1,0) -- (6.5,-2.1) -- (6.5,-2.85);
	\draw[->,orange,opacity=0.6,ultra thick] (p3)+(0.1,0) -- (6.75,-1.4) -- (6.75,-2.85);
	\draw[->,blue!50,opacity=0.6,ultra thick] (p4)+(0.1,0) -- (7.15,-0.15) -- (7.15,-0.4);
	
	\begin{scope}[yshift=-2.8cm]
		\fill[black!5] (0,0) rectangle (5.85,-2.715);
		\node[fill=blue!15,rotate=90,anchor=north east,text width=3.56cm,align=center] at (0,0) {2. Pass};
		\scene
		\draw[CLight] (p1)+(0,-2.8) -- (3.7,-1.7);
		\draw[dashed,red] (p1)++(0,-2.8) -- (4.6,-0.3);
		\draw[dashed,red!50] (p2)++(0,-2.8) -- (4.6,-0.3);
		\draw[CLight] (p3)+(0,-2.8) -- (4.2,-2.2) -- (3.8,-2.5);
		\draw[dashed,red] (p3)++(0,-2.8) -- (4.6,-0.3);
		\draw[CLight,path fading=west] (3.7,-1.7) -- (2,-0.3);
		\draw[CLight,path fading=west] (3.8,-2.5) -- (3.2,-0.6);
		\fill[nee] (2.5,-2.5) circle [radius=1.5pt];
		\fill[nee] (3.9,-2.1) circle [radius=1.5pt];
		\fill[nee] (4.6,-1.4) circle [radius=1.5pt];
		\fill[photon] (3.7,-1.7) circle [radius=1.5pt];
		\fill[photon] (4.2,-2.2) circle [radius=1.5pt];
		\fill[photon] (3.8,-2.5) circle [radius=1.5pt];
	\end{scope}
	
	\begin{scope}[yshift=-5.6cm]
		\fill[black!5] (0,0) rectangle (5.85,-0.8);
		\node[fill=blue!15,rotate=90,anchor=north east,text width=0.87cm,align=center] at (0,0) {
		3.};

		\node[anchor=east] at (4.8,-0.4) {Finalize self emittance};
		\node at (5.1,-0.4) {\Huge$\times$};
	\end{scope}
	\draw[->,orange,opacity=0.6,ultra thick] (6.75,-3.85) -- (6.75,-4.1) -- (5.49,-4.1);
	\draw[->,blue!50,opacity=0.6,ultra thick] (7.7,-1.3) -- (7.7,-6.0) -- (5.49,-6.0);
\end{tikzpicture}
\caption{Algorithm outline: In the first pass a PT is executed and the intermediate results are stored.
In the second pass photons are traced from the non-zero-estimate vertices from pass one.
Finally the contributions from random hits, photons and NEE are weighted to compute the final contribution.}
\label{fig:algorithm} \end{figure}

The basic idea is simple: whenever a path vertex is connected to a light source, we use this very same connection as the first segment of a photon path by creating a virtual light source at the path vertex.
This, however, has several complex implications and a lot of potential for possible modifications.
The outline of the algorithm (Figure \ref{fig:algorithm}) is as follows:
\begin{enumerate}
	\item Trace paths as in \textit{Path Tracing}
	\begin{enumerate}[topsep=0pt]
		\item Store the hit-points (called \textit{NEE vertex})
	\end{enumerate}
	\item NEE and photon tracing
	\begin{enumerate}[topsep=0pt]
		\item Estimate virtual light source density
		\item Compute NEEs
		\item Trace photons and apply contribution directly
		\item \relax[Optional] Trace photons from the light source as usual
	\end{enumerate}
	\item Compute self emittance contributions
\end{enumerate}

\subsection{Trace View Paths}
For tracing paths we use a conventional Path Tracer.
However, it is not possible to compute the results for random hits and NEE immediately.
To calculate the MIS weights, it is necessary to know the photon events which itself depend on the density of NEE vertices (which are being created in this pass).
For this reason the intermediate results for the events must be stored.

\subsection{NEE and Photon Tracing}
Our overall goal is to transform the stored vertices into virtual light sources which emit photons.
To transform the incident differential irradiance (unit \si{\watt\per\square\meter}), which is available through NEE, into a flux $\Phi$ (unit \si{\watt}) we need the density of source vertices.
Effectively, we need to know the area $A$ of the virtual light source to integrate the incoming irradiance:
\begin{equation}
	\Phi = \text{d}E \cdot \text{d}A = \frac{\text{d}E}{\rho}. \label{eq:flux}
\end{equation}
where $\rho$ is the density of NEE vertices at the current virtual light source.
We can insert $\text{d}A=1/\rho$ because
the area, which is associated with each vertex, depends on this density:
If there are more vertices, each one represents a smaller part of the surface area.

Note that it is not important how the vertices were generated.
It only matters how many vertices are stored in a local area to turn each of them into an unbiased\footnote{Unbiased, if we know the true density, which is not the case} emitter.
Especially, the past (sampling events, Russian Roulette, ...) of the view path vertices is not important.

So, in step (a) the density of vertices must be computed at each vertex.
This can be achieved with k-nearest neighbor searches or and additional data structure.
We use a sparse octree to integrate density over regions and time, as will be detailed in Section \ref{sec:density}.
The advantage of the octree is a higher performance and less noise in the estimates.

Now (step (b)), we can compute a next event estimation for each of the stored vertices and compute its contribution
\begin{align}
	L_E &= w_{E,k}\cdot \tau(\nu_k) \cdot f(\nu_k, \mvec{d})\cdot \text{d}E
\end{align}with
\begin{align*}
	\mvec{d} &= \text{Direction of NEE}\\
	\text{d}E &= \text{Differential irradiance of NEE}\\
	\nu_k &= \text{Vertex with index $k$ (camera has index 0)}\\
	\tau(\nu) &= \text{Path throughput from MC sampling } \prod_{i=0}^k \frac{f_i}{p_i}\\
	f(\nu,\mvec{d}) &= \text{Bidirectional Scattering Distribution Function (BSDF)}\\
	w_{E,k} &= \text{MIS weight (see Section \ref{sec:mis}).}
\end{align*}
It must be added to the pixel, in which the path originated.
The details to compute the MIS weight $w_{E,k}$ are given in the next section.

Photons can be traced (step (c)) by starting a new random walk at the current vertex, after applying Equation \eqref{eq:flux}.
The first sampling event is based on the NEE connection direction $\mvec{d}$ as incident direction and the vertex's BSDF.
It is not necessary to store photons, because we can invert the search for merge events.
Instead of searching for photons around the view path vertices, we can as well search for view path vertices around photons, which produces identical results.
Therefore, the NEE vertex storage must be some kind of spatial data structure to support neighborhood searches.

Then the contribution of a photon $\Phi_i$ at $\nu_i$ from direction $\mvec{d}_i$ for a found vertex $\nu_k$ is
\begin{equation}
	L_{P} = w_{P,k,i}\cdot \tau(\nu_k) \cdot f(\nu_k, \mvec{d}_i)\cdot \Phi_i \cdot K(\lVert \nu_k - \nu_i \rVert)
\end{equation}
where details on the weight will follow in the next section, again.
The kernel $K$ can be the uniform kernel $1/\pi r^2$ for the search radius $r$ or any other kernel used in photon mapping.

Optionally, it is possible to trace photons from the light source (step (d)) and add their contributions the same way as the photons from NEE vertices.
Only the MIS weights must be adapted accordingly.
We will show in section \ref{sec:modifications} that adding those photons complements NEB where it is weakest.

\subsection{Compute Self Emittance Contributions}

Since it was not possible before step (2.a) to compute the MIS weights properly, we needed to store random hits of light source in pass~1.
Now, we can iterate over the stored events and compute the contributions
\begin{equation}
	L_e = w_e \cdot \tau(\nu_k) \cdot L_e(\nu_k)
\end{equation}

\section{MIS Weights}
\label{sec:mis}

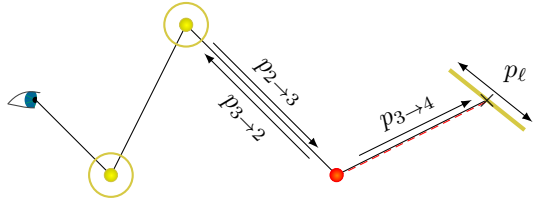
\begin{figure} \centering
\begin{tikzpicture}
	\begin{scope}[scale=1.5]\draweyex{0,0};\end{scope}
	\draw (0,0) -- (1,-1) -- (2,1) -- (4,-1) -- (6,0);
	\draw[red,densely dashed] (4,-1.02) -- (6,-0.02);
	\draw[ultra thick,CLight] (6.5,-0.4) -- (5.5,0.4);
	\draw[CLight,thick] (1,-1) circle [radius=0.29];
	\fill[photon] (1,-1) circle [radius=2.5pt];
	\draw[CLight,thick] (2,1) circle [radius=0.29];
	\fill[photon] (2,1) circle [radius=2.5pt];
	\fill[nee] (4,-1) circle [radius=2.5pt];
	\node at (6,0) {$\times$};
	\draw[<->] (45:0.15)+(6.5,-0.4) -- +(5.5,0.4) node[pos=0.5,anchor=south west] {$p_\ell$};
	\draw[->] (4.3,-0.74) -- (5.8,0.0) node[pos=0.5,anchor=south,sloped] {$p_{3\rightarrow4}$};
	\draw[->] (2.36,0.76) -- (3.76,-0.64) node[pos=0.5,anchor=south,sloped] {$p_{2\rightarrow3}$};
	\draw[<-] (2.24,0.62) -- (3.64,-0.78) node[pos=0.5,anchor=north,sloped] {$p_{3\rightarrow2}$};
\end{tikzpicture}
\caption{MIS-weight computation between several events along the same path.
The shown example path has length $\ell=4$.}
\label{fig:mis} \end{figure}

In our basic algorithm there are three types of events which must be weighted against each other.
When conventional photons are traced too, there is one additional event type.
Each path of length $\ell$ has a single random hit event, one NEE event if $\ell\geq2$ and $\max(0,\ell-2)$ photon merge events as depicted in Figure \ref{fig:mis}.
Optionally, there are $\max(0,\ell-1)$ merges with usual photons.

In general, MIS-weights can be computed with the balance heuristic \cite{veach_optimally_1995} which reads
\begin{equation}
	w_{\!j} = \frac{n_{\!j} p_{\!j}^*}{\sum_i n_i p_i^*}
	\label{eq:misweight}
\end{equation}
where $p^*$ are the path sampling densities of the different events, with respect to the area measure, and $n$ are the number of samples drawn with the respective sampler.

For random hits we have $n_e=1$ and
\begin{equation}
	p_{e}^* = \prod_{i=0}^{\ell-1} p_{i\rightarrow i+1} \label{eq:rhpdf}
\end{equation}
for the path density up to the vertex $k$.
The \textit{Probability Density Functions} (PDFs) $p_{i\rightarrow i+1}=p_i\cdot \cos\theta / d^2$ (unit \si{\per\square\meter}) describe the probability to reach vertex $\nu_{i+1}$ via random sampling of the BSDF at vertex $\nu_i$ with the PDF $p_i$.

The path density for NEEs is defined as
\begin{equation}
	p_{E}^* = p_\ell \prod_{i=0}^{\ell-2} p_{i\rightarrow i+1} \label{eq:neepdf}
\end{equation}
where $p_\ell$ is the sampling density per area to sample the specific point on the light source when attempting a random connection.
The number of NEE samples is usually but not necessarily $n_E=1$.

Next, we need the PDF for the new photon samplers.
It is closely related to the NEE PDF in Equation \eqref{eq:neepdf} because each photon path starts with a next event estimation and therefore has to use $p_\ell$.
From there, a random walk in backward direction is performed up to the merge vertex $\nu_k$.
Further, it consists of another random walk beginning at the observer.
Finally, we need the chance for a successful merge $\rho_{\ell-1} \cdot \pi r^2 / n_T$ with $n_T$ being the total number of photon path starting points, i.e. the number of stored NEE vertices.
Together this gives
\begin{equation}
	p_{P,k}^* = p_\ell \cdot \frac{\rho_{\ell-1} \cdot \pi r^2}{n_T} \cdot \prod_{i=k}^{\ell-2} p_{i+1\rightarrow i} \cdot \prod_{i=0}^{k-1} p_{i\rightarrow i+1} \label{eq:photonpdf}
\end{equation}
and $n_{P} = n_T$ for the sampler count because we reuse photons from all paths.
The above merge chance is derived as following:
First, it must be proportional to the size of our search region $\pi r^2$.
The larger the search region, the larger the chance to find something.
Second, we need the probability density per area to start the respective photon random walk.
The density $\rho_{\ell-1}$ is the one we computed in step (2.a).
It is the total number of events per area at this start vertex.
Thus, $\rho_{\ell-1} / n_T$ is the PDF per area of a single sample.
Both together give the chance to find the chosen photon sub-path.

Finally, we have
\begin{equation}
	p_{LP,k}^* = p_\ell \cdot \pi r^2 \cdot \prod_{i=k}^{\ell-1} p_{i+1\rightarrow i} \cdot \prod_{i=0}^{k-1} p_{i\rightarrow i+1} \label{eq:stdphotonpdf}
\end{equation}
for the conventional photons \cite{georgiev_light_2012,hachisuka_path_2012}.
The number of samples is the number of additional photon paths $n_{LP}$ due to global reuse of photons or $n_{LP} = 0$ if disabled.

Plugging Equations \eqref{eq:rhpdf} to \eqref{eq:stdphotonpdf} and the sampler counts into the balance heuristic \eqref{eq:misweight} gives us the searched weights:
\begin{align*}
	w_{e}\! = \!\frac{p_{e}^*}{p_\text{sum}} &&
	w_{E}\! = \!\frac{n_{E} \cdot p_{E}^*}{p_\text{sum}} &&
	w_{P,k}\! = \!\frac{n_P\cdot p_{P,k}^*}{p_\text{sum}} &&
	w_{LP,k}\! = \!\frac{n_{LP}\cdot p_{LP,k}^*}{p_\text{sum}}
\end{align*}%
\begin{equation}
	\text{with } p_\text{sum} = p_{e}^* + n_{E} \cdot p_{E}^* +
		n_P \sum_{k=1}^{\ell-2} p_{P,k}^* + 
		n_{LP} \sum_{k=1}^{\ell-1} p_{LP,k}^* \label{eq:misw}
\end{equation}

\section{Density Estimation}
\label{sec:density}

An important point for the correctness and performance of the NEB operator is the estimate of the density $\rho$ (required in Equation \eqref{eq:flux}).
Density estimation is a well explored research area for which we refer to the book of Silverman \cite{silverman_density_1986} and the survey of Sheather \cite{sheather_density_2004}.
Unfortunately, the density estimation becomes the bottleneck of the algorithm fast and the choice of the data structure is very important.

Our first approach was to use a hash-grid to query the number of photons with a predefined search radius.
This is referred to as na{\"i}ve estimator in Silverman's book and has a fundamental drawback:
Its bias gets very large if the distribution of NEE vertices is irregular.
In low density regions (less than one vertex per search region) the estimate will always find the query point, but likely no others.
Therefore, it may overestimate the density by an unbounded factor.
On the other hand, in high density regions it will blur the density function more than necessary.

Our second approach was to use a kd-tree \cite{bentley_multidimensional_1975} to estimate the density with the \textit{k-nearest neighbor} approach.
The nearest neighbor approach scales much better with irregular densities.
We found that the bias was acceptable with a $k\geq 4$ neighbors.
However, the kd-tree maintenance and query time dominated our rendering time vastly.
We did not try to use faster builder implementations like the ParKD method from Choi et al. \cite{choi_parallel_2010}, because they would not help to improve the query time performance.

\subsection{The Density Octree}
\label{sec:densityoctree}

To improve performance, we implemented a dedicated data structure for the density estimate.
Other than the hash-grid or the kd-tree, this new data structure is not able to find the actual vertices because we store particle counts explicitly.
Therefore, it achieves a speedup of $3000\times$ opposed to the kd-tree while having a comparable bias.

\begin{figure} \centering
\begin{tikzpicture}
	\begin{scope}[scale=0.75]
		\fill[black!5] (0,2) rectangle +(2,2);
		\fill[black!5] (2,0) rectangle +(2,2);
		\fill[black!10] (0,0) rectangle +(2,1);
		\fill[black!10] (0,1) rectangle +(1,1);
		\fill[black!10] (2,2) rectangle +(2,1);
		\fill[black!10] (2,3) rectangle +(1,1);
		\fill[black!20] (1,1) rectangle +(1,1);
	
		\draw (0,0) rectangle (4,4);
		\draw (0,2) -- (4,2) (2,0) -- (2,4);
		\draw (0,1) -- (2,1) (1,0) -- (1,2);
		\draw (2,3) -- (4,3) (3,2) -- (3,4);
		\draw (1,1.5) -- (2,1.5) (1.5,1) -- (1.5,2);
		
		\fill[nee] (0.3,3.7) circle [radius=2pt];
		\fill[nee] (0.6,2.2) circle [radius=2pt];
		\fill[nee] (1.7,1.8) circle [radius=2pt];
		\fill[nee] (1.3,1.9) circle [radius=2pt];
		\fill[nee] (1.1,1.6) circle [radius=2pt];
		\fill[nee] (1.2,1.3) circle [radius=2pt];
		\fill[nee] (1.4,1.2) circle [radius=2pt];
		\fill[nee] (1.6,1.4) circle [radius=2pt];
		\fill[nee] (0.4,1.5) circle [radius=2pt];
		\fill[nee] (0.3,0.5) circle [radius=2pt];
		\fill[nee] (0.7,0.7) circle [radius=2pt];
		\fill[nee] (1.9,0.4) circle [radius=2pt];
		\fill[nee] (2.4,1.1) circle [radius=2pt];
		\fill[nee] (2.5,2.4) circle [radius=2pt];
		\fill[nee] (2.7,2.6) circle [radius=2pt];
		\fill[nee] (2.4,3.1) circle [radius=2pt];
		\fill[nee] (3.3,2.2) circle [radius=2pt];
	\end{scope}
	
	\begin{scope}[xshift=5cm]
		\draw[densely dotted] (0,0) -- (1,1) -- (1,3) (1,1) -- (3,1);
		\node[anchor=east] at (1,1) {$\mvec{b}_{100}$};
		
		\draw[fill=blue!20,opacity=0.4] (0,0.9) -- (0.8,0) -- (2.6,0.6) -- (3,1.6) -- (1.2,3) -- (0.85,2.85) -- cycle;
		\draw (0,0) rectangle (2,2);
		\draw (2,2) -- (3,3) (0,2) -- (1,3) (2,0) -- (3,1);
		\draw (1,3) -- (3,3) -- (3,1);
		
		\fill (0,0.9) circle [radius=1.5pt];
		\fill (0.8,0) circle [radius=1.5pt];
		\fill (0.85,2.85) circle [radius=1.5pt];
		\fill (1.2,3) circle [radius=1.5pt];
		\fill (3,1.6) circle [radius=1.5pt];
		\fill (2.6,0.6) circle [radius=1.5pt];
		
		\draw[->] (1.2,1.2) -- (1.35,1.7) node[anchor=west] {$\mvec{n}$};
		\fill[nee] (1.2,1.2) circle [radius=2pt];
		\node[anchor=north] at (1.2,1.2) {$\mvec{p}$};
		
		\node[anchor=east] at (0,0) {$\mvec{b}_{000}$};
		\node[anchor=east] at (0,2) {$\mvec{b}_{010}$};
		\node[anchor=east] at (1,3) {$\mvec{b}_{110}$};
		\node[anchor=west] at (3,3) {$\mvec{b}_{111}$};
		\node[anchor=west] at (3,1) {$\mvec{b}_{101}$};
		\node[anchor=west] at (2,0) {$\mvec{b}_{001}$};
		\node[anchor=west] at (2,2) {$\mvec{b}_{011}$};
	\end{scope}
\end{tikzpicture}
\caption{Concepts of the density octree.
Left: a sparse quadtree with a uniform count of particles per cell.
Right: one possible case for the intersection area between plane and cube.}
\label{fig:densityoctree} \end{figure}
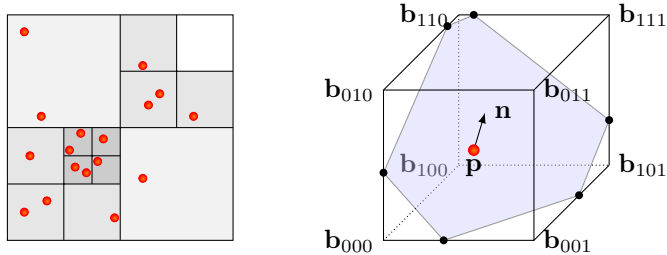

The idea is to use a sparse octree which stores an atomic counter per cell.
Whenever a vertex is created, it increases the counter of the cell in which it lies by one.
If the number is greater than some predefined threshold, the cell is split into eight new cells.
The resulting tree has large cells in low density regions and small ones in high density regions, like the kd-tree.
An example quadtree (2D octree) is shown in Figure \ref{fig:densityoctree}.

\subsection{Splitting}

If splitting a cell, we need to initialize the eight new cells.
Unfortunately, the distribution of points which filled the cell is not known anymore and we need to make a guess.
Without further knowledge we can only distribute the counter uniformly to all child cells.

However, dividing the counter by eight, the number of children, systematically produces an underestimated initialization.
The reason is that a 2D surface only intersects expectedly four of the eight children.
Therefore, using the parent counter divided by four as initialization value turned out to be less biased in practice.
The too large values in cells without an surface intersection do not matter, since they are never queried.

\subsection{Queries}

When a query is made, the count $c$ in the respective cell must be converted to a density.
Assuming locally flat surfaces and that the current surface is the only one inside the cell, the intersection area between the cell's bounding box $\mathcal{B}$ and the plane $\mathcal{P}$ can be calculated.
Thereby, the plane is defined by the surface position $\mvec{p}$ and normal $\mvec{n}$.
Thus, our density estimate is
\begin{equation}
	\rho = \frac{c}{|\mathcal{B} \cap \mathcal{P}|} = \frac{c}{A}, \label{eq:octdensity}
\end{equation}
where the intersection area is computed with
\begin{align}
	A &= \begin{cases}
		\mvec{s}_x \cdot \mvec{s}_y, & \mvec{n}_x\! =\! \mvec{n}_y\! =\! 0\\
		\mvec{s}_x \left|\frac{1}{\mvec{n}_y \mvec{n}_z} \sum\limits_{i=0}^3 \epsilon_i \max\left(0, \langle\mvec{n},\mvec{p}\rangle - \langle\mvec{n},\mvec{b}_i\rangle \right) \right|, & \mvec{n}_x\! =\! 0\\
		\left|\frac{1}{2\mvec{n}_x \mvec{n}_y \mvec{n}_z} \sum\limits_{i=0}^7 \epsilon_i \max\left(0, \langle\mvec{n},\mvec{p}\rangle - \langle\mvec{n},\mvec{b}_i\rangle \right)^2 \right|
	\end{cases} \label{eq:area}\\
	\mvec{s} &= \mvec{b}_{111} - \mvec{b}_{000}\nonumber\\
	\epsilon_i &= \begin{cases}
		\phantom{-}1, & i\in\{000, 011, 101, 110\}\nonumber\\
		-1, & i\in\{100, 010, 001, 111\}
	\end{cases}.
\end{align}
Here, $x$, $y$ and $z$ are the indices of the dimensions, $\mvec{s}$ is the size of the box and $\epsilon_{i}$ is the parity of the eight vertices.
In the first case, two of the dimensions are zero.
W.l.o.g. only $\mvec{n}_z \neq 0$ is shown.
The terms for the other two dimensions are defined analogously.
Similarly, case two shows the situation where one dimension of $\mvec{n}$ is zero.
Again, there are three analogous terms for all dimensions.
The third case applies if no dimension is zero.
A derivation can be found in Appendix~\ref{app:area}.

\begin{figure} \input{figures/densityoctreevis} \end{figure}

An advantage of the dedicated structure is that the density can be integrated over time.
Therefore, the split threshold must be increased proportionally to the iteration count and Equation \eqref{eq:octdensity} must be divided by the number of iterations.
This reduces noise and increases the independence between the current sample set and the density estimate.
Figure \ref{fig:densityvis} visualizes the query results and shows a fast convergence of the estimate.
After four iterations the octree already has significantly less noise than the kd-tree-based search.
Unfortunately, there are small dark points (see image on the right) which are caused by floating-point precision issues.
These can cause too bright photons in the image which are often hidden by the MIS.

\subsection{Memory Layout Details}
We store the eight counters of the sibling cells in a consecutive sequence.
This allows us to use a single pointer in a parent to address all its children.
Moreover, it is possible to encode the counter and the child index in the same integer to save memory.
If the stored value is negative, its absolute value is the index of the first child node.
Otherwise it is a counter.

The aforementioned split threshold is set to four in our implementation.
Using larger numbers introduces too much bias, because the blurring area increases and the planar surface assumption is less valid.
Using smaller numbers increases the memory consumption and leads to more noise.

The necessary memory depends on two factors only:
the expected number of NEE vertices divided by the split threshold.
Particularly, it is independent of the scene.
For typical setups the density octree requires less than 50 MB (often less than 10 MB suffice).

For more details on the lock-free implementation, we provide the code in Appendix~\ref{app:octreeimpl}.

\section{Modifications}
\label{sec:modifications}

Having a robust density estimate, all tools for NEB are given.
In this section, we evaluate simple modifications to improve the performance or robustness of the basic algorithm

\begin{figure} \includegraphics[width=0.495\linewidth]{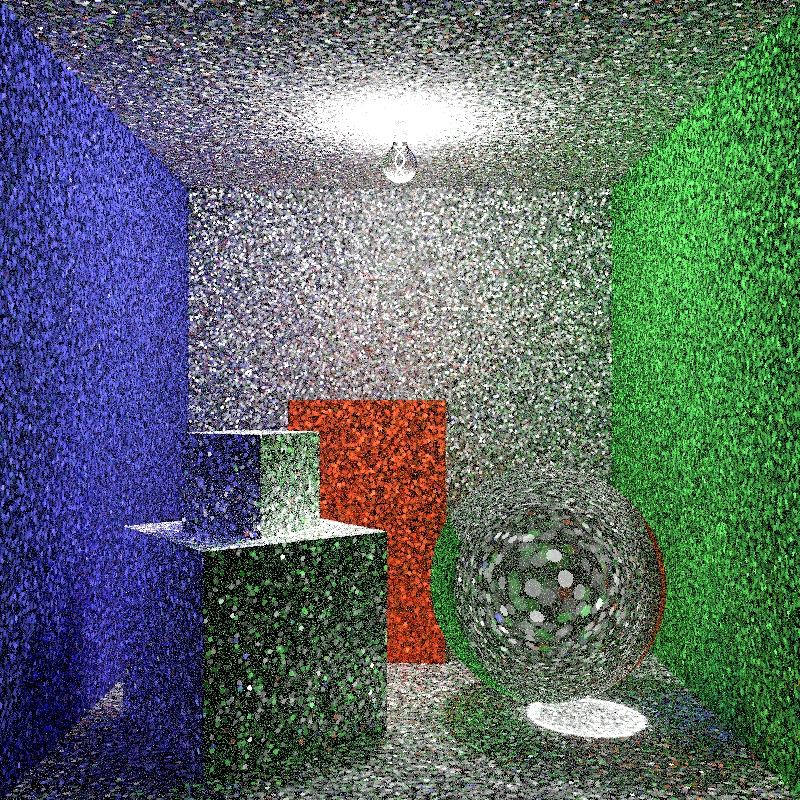}~%
\includegraphics[width=0.495\linewidth]{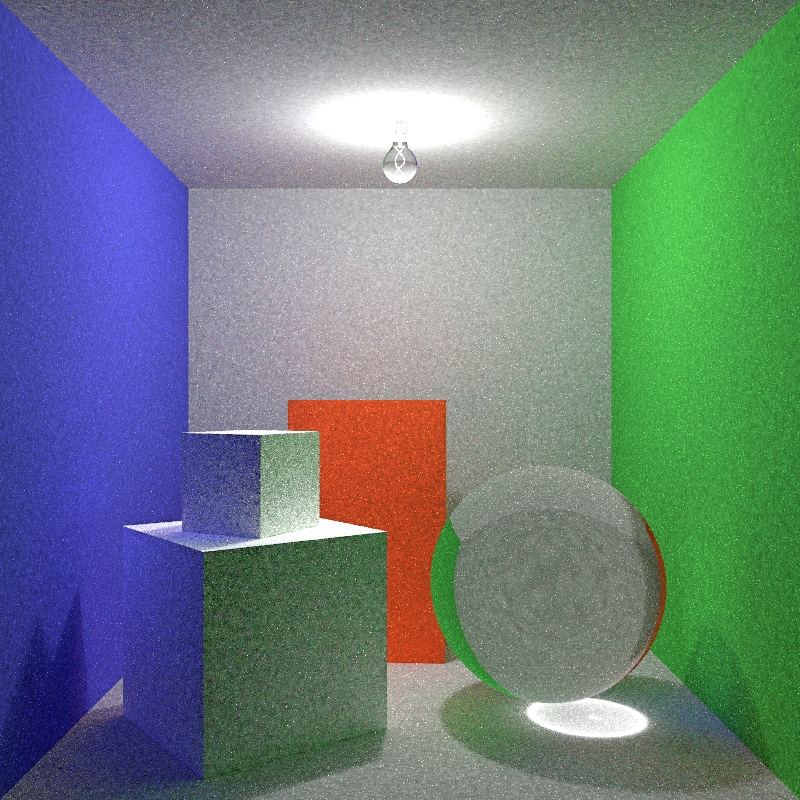}\\[1pt]
\begin{tabularx}{\linewidth}{CC}
	NEB, 68spp & NEB + LP, 14spp
\end{tabularx}

\caption{Equal time comparison (5 \si{\minute}) without and with additional photons (Option~2.c., Section \ref{sec:lightphotons}) in a light bulb scenario.}
\label{fig:secondarynees}
 \end{figure}
\begin{figure*} \centering
Single NEE\\[2pt]
\begin{tabularx}{\linewidth}{CCCCC}
  $\ell=2$ & $\ell=3$ & $\ell=4$ & $\ell=5$ & $\ell\in[2,5]$
\end{tabularx}
\includegraphics[width=0.196\linewidth]{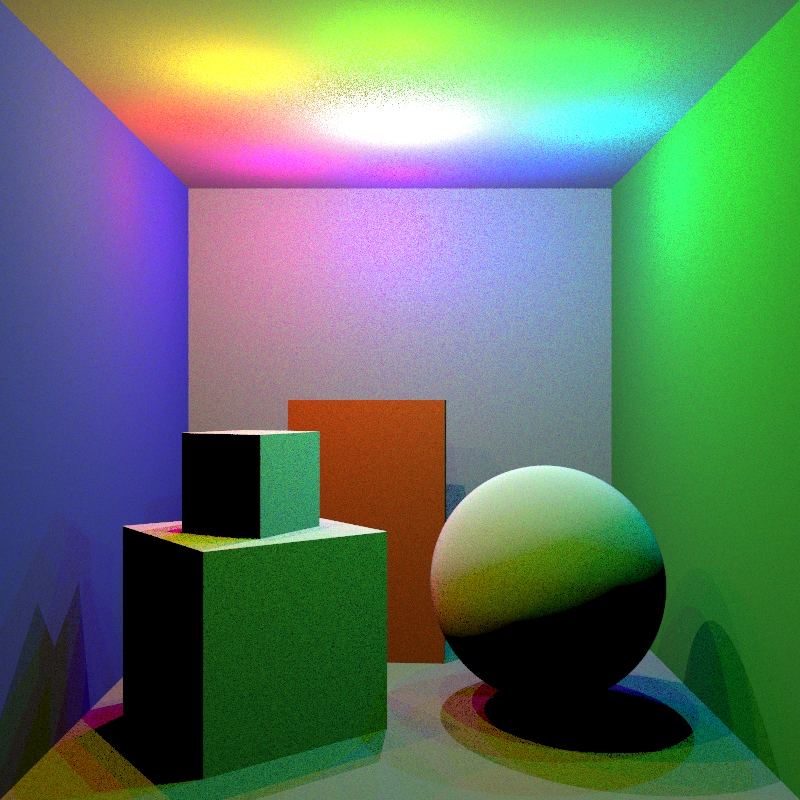}~%
\includegraphics[width=0.196\linewidth]{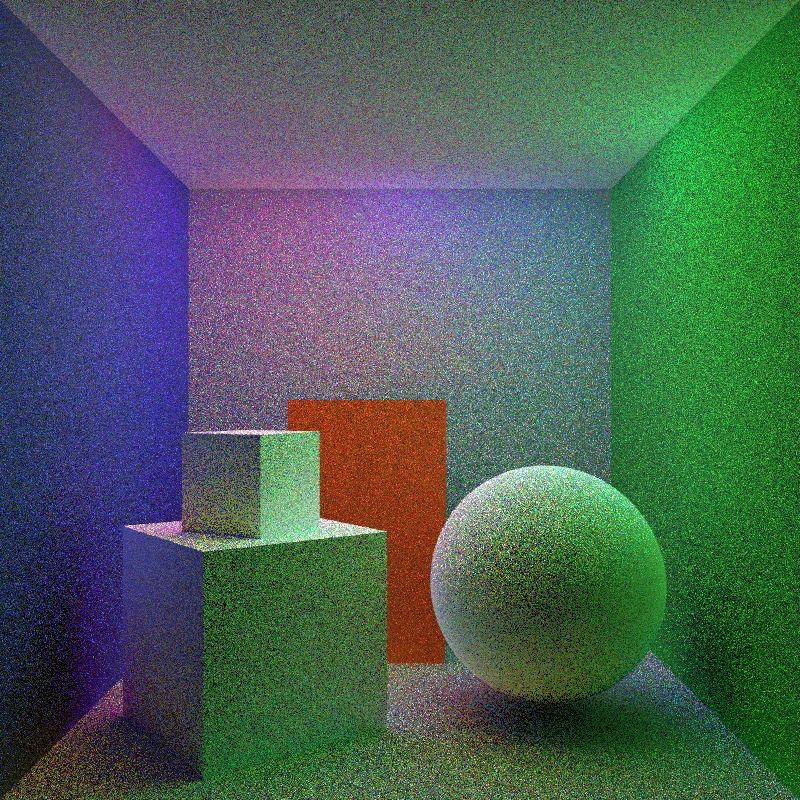}~%
\includegraphics[width=0.196\linewidth]{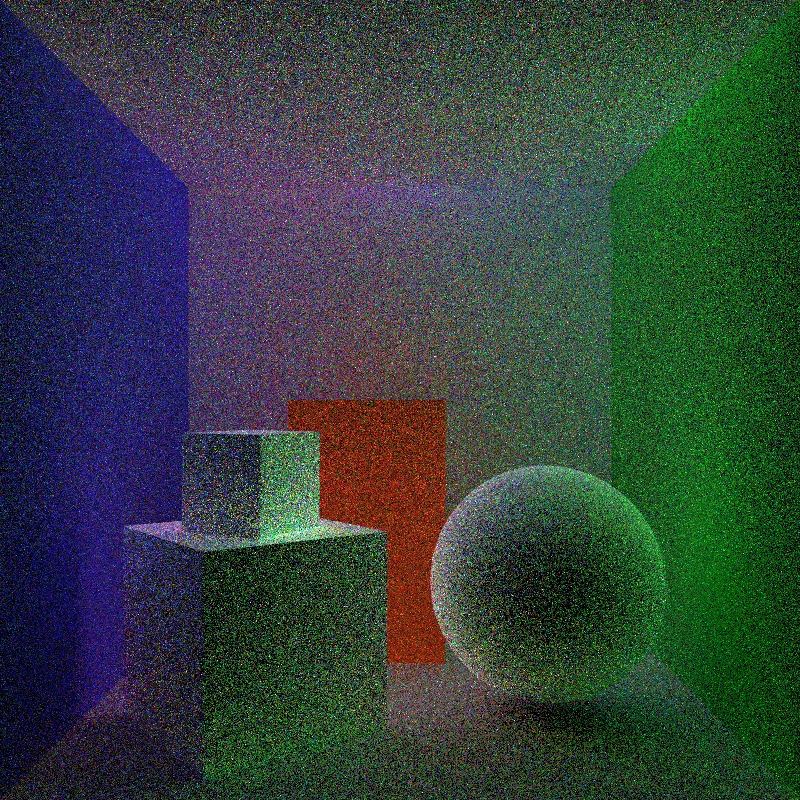}~%
\includegraphics[width=0.196\linewidth]{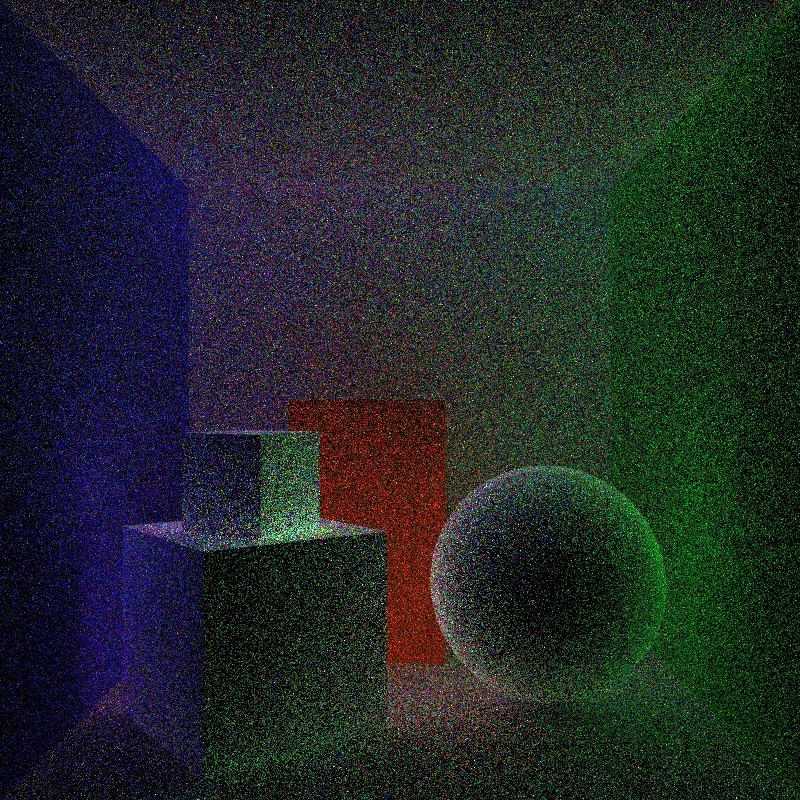}~%
\includegraphics[width=0.196\linewidth]{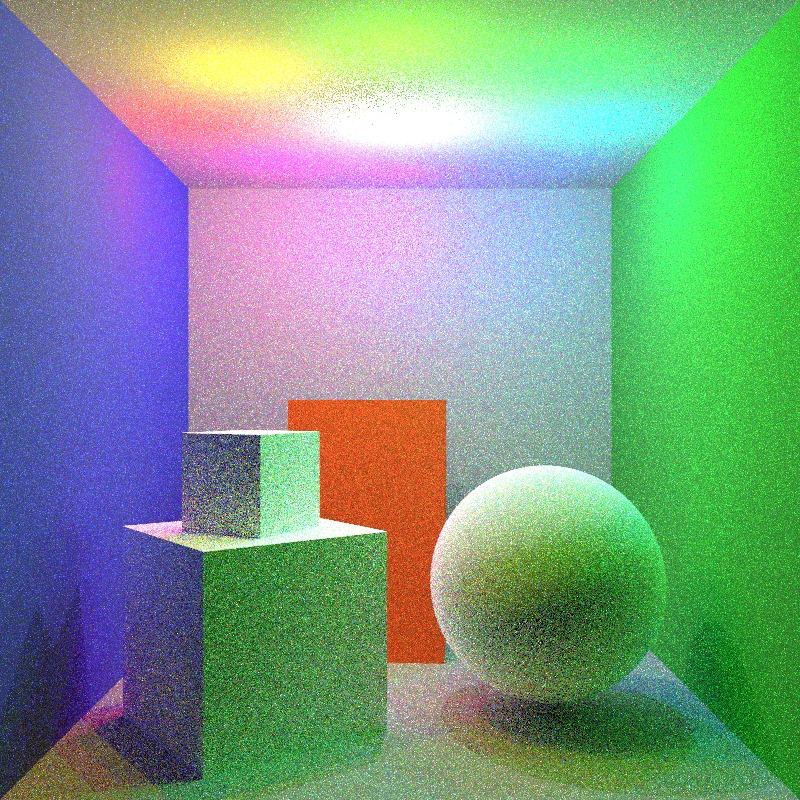}\\
\begin{tabularx}{1.02\linewidth}{CCCCC}
  29~spp, RMSE~0.198 & 19~spp, RMSE~0.201 & 16~spp, RMSE~0.176 & 14~spp, RMSE~0.152 & 14~spp, RMSE~0.457
\end{tabularx}\\[8pt]

Merged NEE\\
\includegraphics[width=0.198\linewidth]{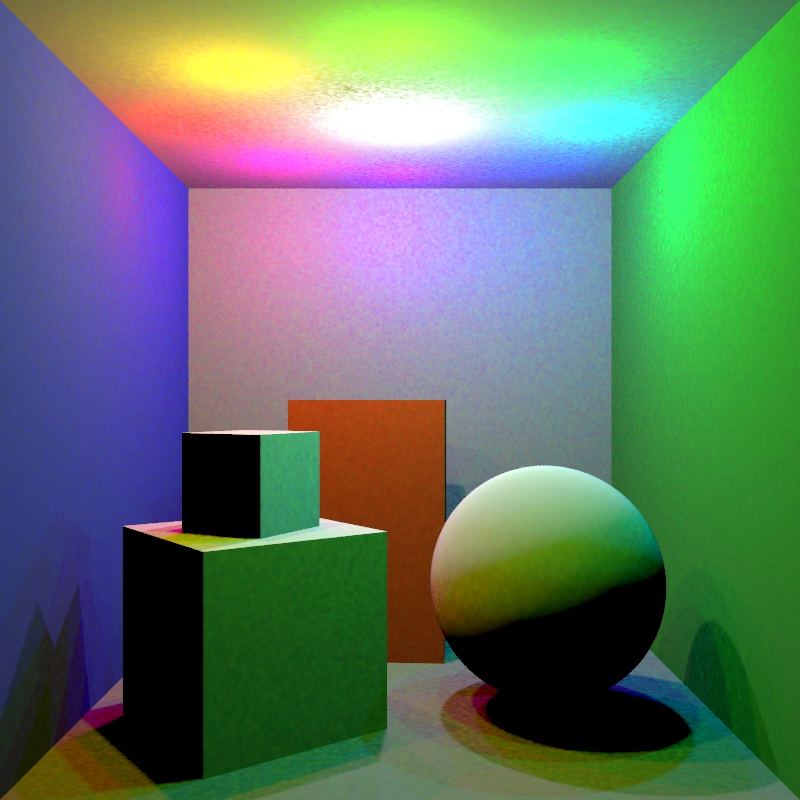}~%
\includegraphics[width=0.198\linewidth]{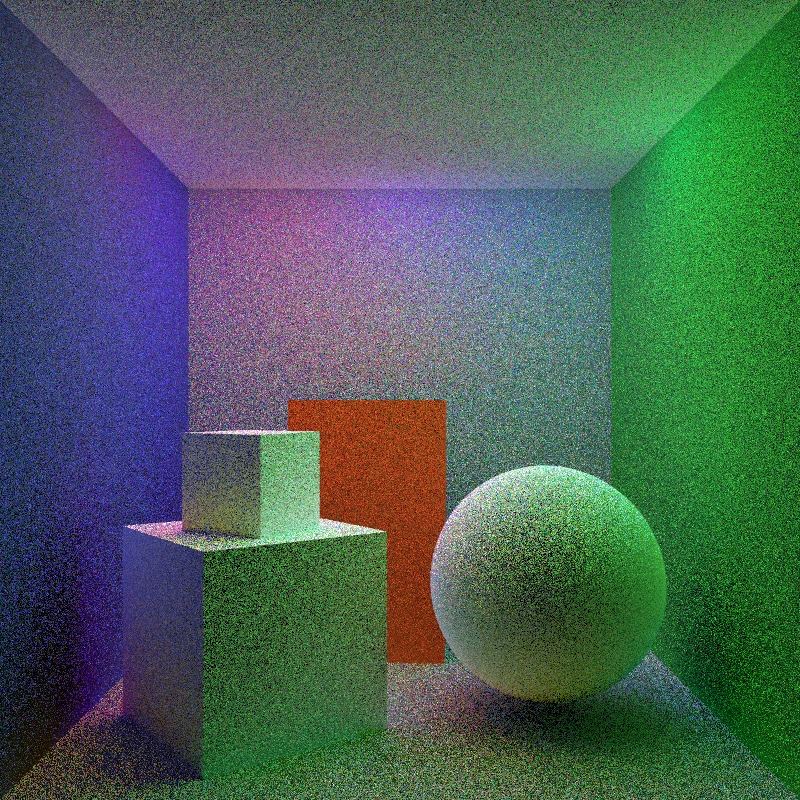}~%
\includegraphics[width=0.198\linewidth]{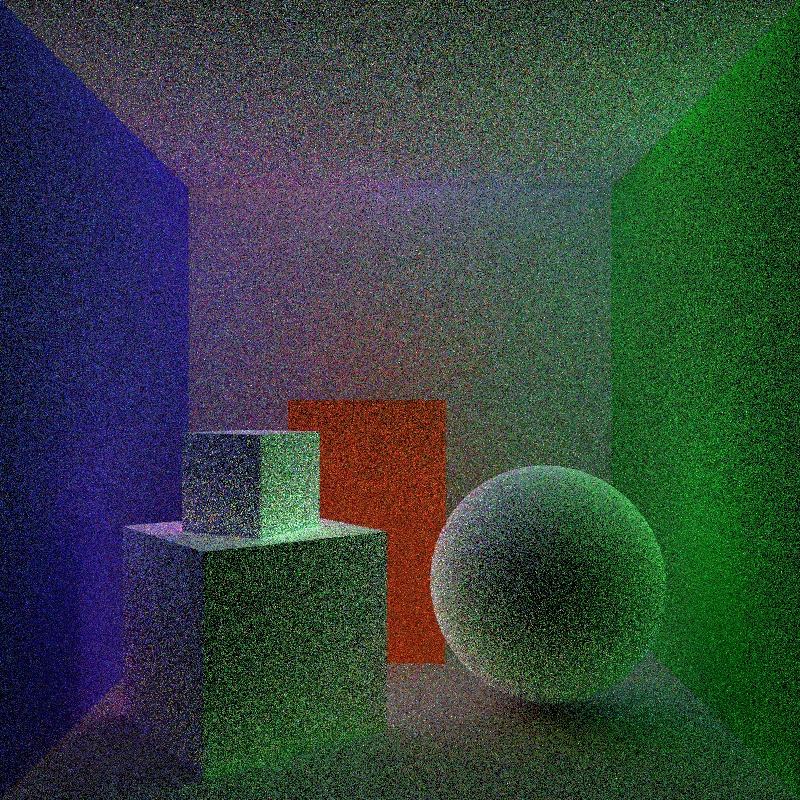}~%
\includegraphics[width=0.198\linewidth]{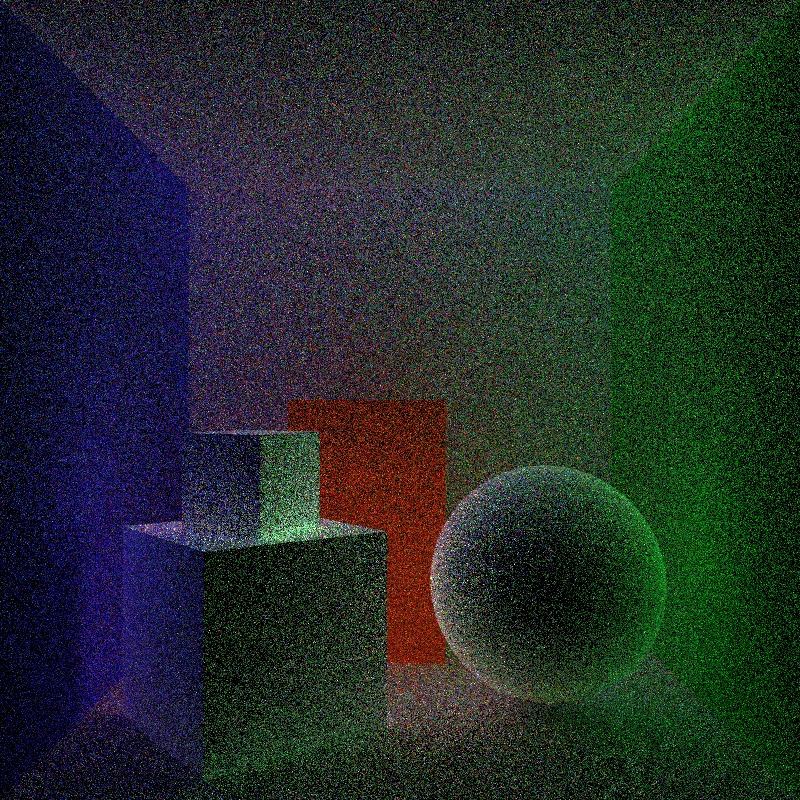}~%
\includegraphics[width=0.198\linewidth]{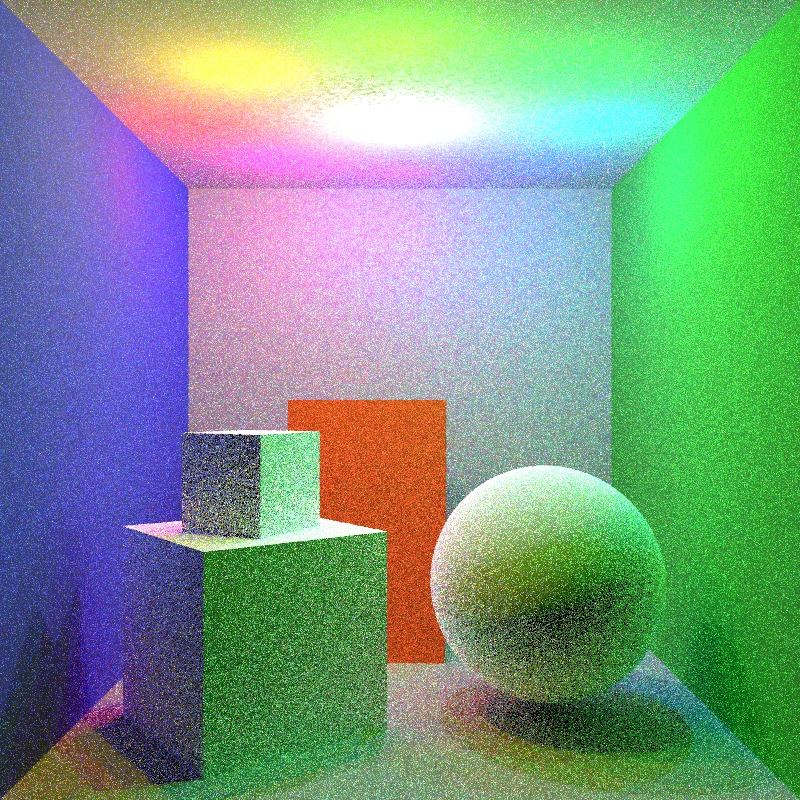}\\
\begin{tabularx}{1.02\linewidth}{CCCCC}
  16~spp, RMSE~0.104 & 11~spp, RMSE~0.178 & 12~spp, RMSE~0.141 & 12~spp, RMSE~0.113 & 5~spp, RMSE~0.455
\end{tabularx}

\caption{Equal time comparison (1 \si{\minute}) of Path Tracing with and without NEE recycling.}
\label{fig:neerecycling}
 \end{figure*}

\subsection{Conventional Light Photons}
\label{sec:lightphotons}

While the basic algorithm is very strong in scenarios like the teapot in a stadium (Figure \ref{fig:teaser}), it fails when the caustic throwing object is much closer to the light source than to the receiver.
Consider the example of a light bulb -- an emitter inside a glass ball.
Only NEEs on the surface of the bulb produce contributing photons, but only very few paths randomly hit the comparable small bulb.
All other vertices in the scenes have no NEE contribution, because the glass ball is blocking the connection to the light source.
Therefore, the number of useful NEEs and photons are both very small, leading to high variance results.

We observed that the failure cases of NEB occur in situations where the conventional photon tracing, which starts at light sources, is strong.
Hence, combining NEE photons and light photons by the means of MIS promises a more robust algorithm.
In Figure~\ref{fig:secondarynees} we demonstrate the effectiveness of this combination.
The renderer becomes much more effective with respect to time although its iteration count decreases.
The reason for both is that many more photons are found and merged at each position.
Besides the additional tracing of photons, this requires more evaluations of the BSDF and the MIS weights.

\subsection{NEE Recycling}
\label{sec:neemerges}

Since we already store the NEE vertices in a search data structure, it seems reasonable to share the results of NEE events.
Therefore, it is necessary to store the NEE information along with the vertices and to query those with a neighborhood search.
Then, all available NEEs at one vertex must be averaged and the effective count of NEEs $n_E$ in Equation \eqref{eq:misw} increases to the number of found events.

In Figure \ref{fig:neerecycling} we show an experiment with and without NEE reuse to judge the effectiveness of the proposed modification.
Enabling the merges is clearly slower due to the range query and the additional evaluations of BSDFs.
For the range queries we used a hash grid with a fixed query radius.
Despite the lower number of samples, the noise level (\textit{Root Mean Squared Error}) is slightly better when reusing the NEEs.
However, the effectiveness decreases with path length and gets worse than usual PT for practical path lengths.
Repeated experiments with different merge radii had the same outcome.
The reason for the low effectiveness is that the noise in indirect lighting is dominated by the random walk and not the NEE.

Concluding, the idea of reusing the NEE events sounds promising, but does not pay off in this form.
Therefore, we used only the one primary NEE without this modification all other experiments.

\section{Comparison to Other Methods}
\label{sec:comparison}

The next event backtracking operator has clear strengths and weaknesses.
It is strong whenever NEE is more likely than other events on the path.
This is the case for small or distant light sources like in Figure \ref{fig:teaser}.
Additionally, it scales well with many lights, if contribution-sensitive NEEs are used.
It is weak, whenever the vertex density is much smaller than the light contribution.
In this situation, combining NEB with light photons (+LP) can alleviate this problem.

\begin{figure} \centering%
\setlength\tabcolsep{0.5pt}%
\begin{tabularx}{\linewidth}{Ccccc}
& \textsc{Caustics} & \textsc{SDS} & \textsc{Mirrors} & \textsc{Reflector} \\
\rotatebox{90}{\hspace{1.8em}Reference} &
  \includegraphics[width=0.237\linewidth]{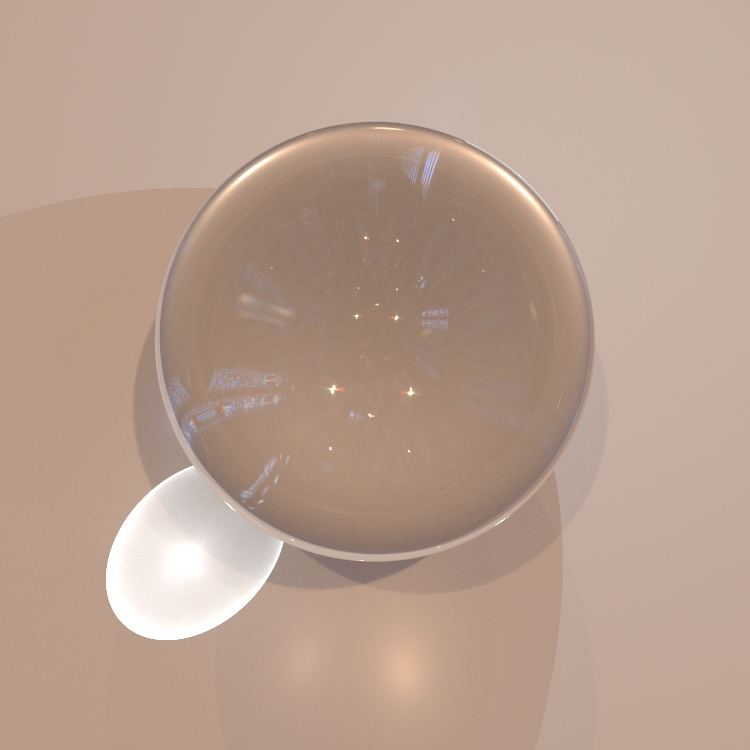} &
  \includegraphics[width=0.237\linewidth]{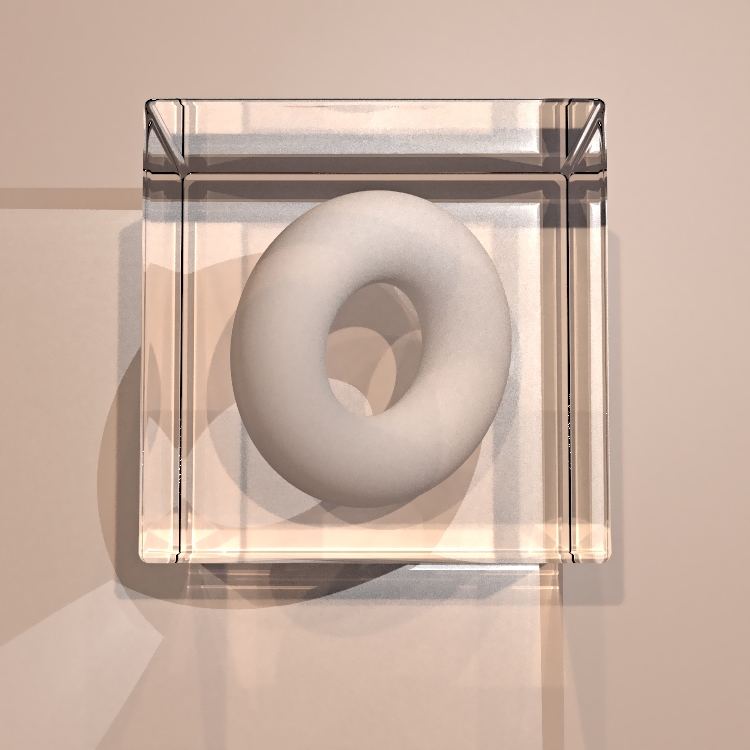} &
  \includegraphics[width=0.237\linewidth]{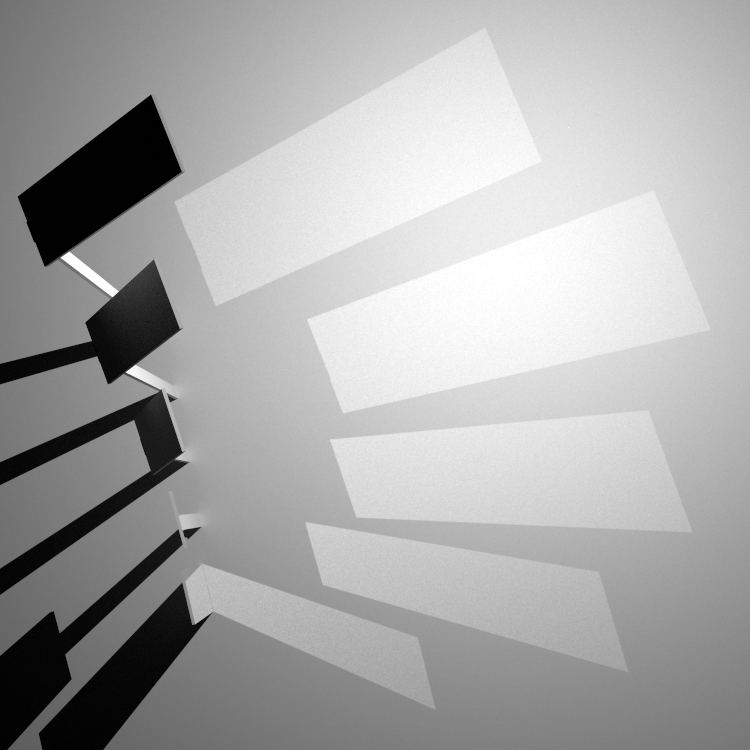} &
  \includegraphics[width=0.237\linewidth]{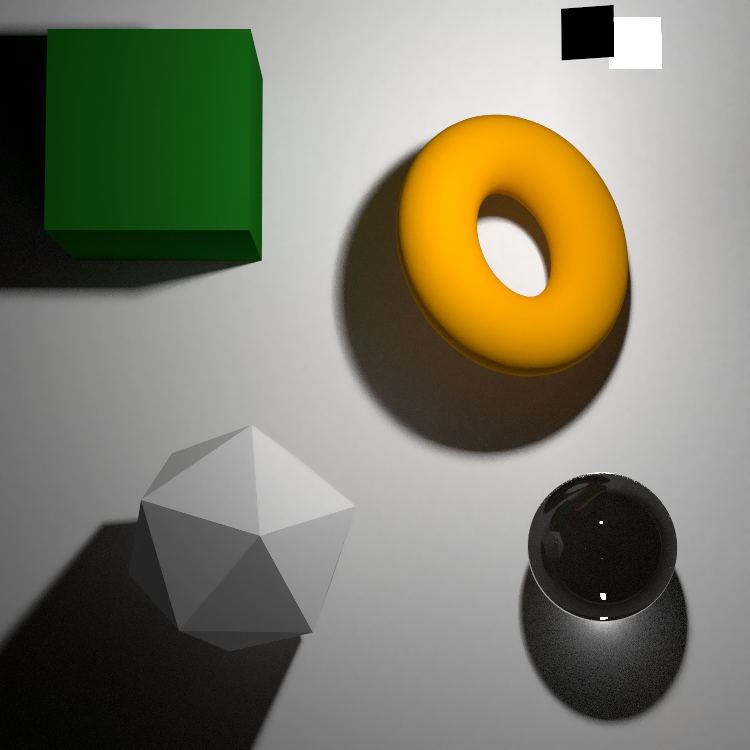}\\
\rotatebox{90}{\hspace{2.9em}NEB} &
  \includegraphics[width=0.237\linewidth]{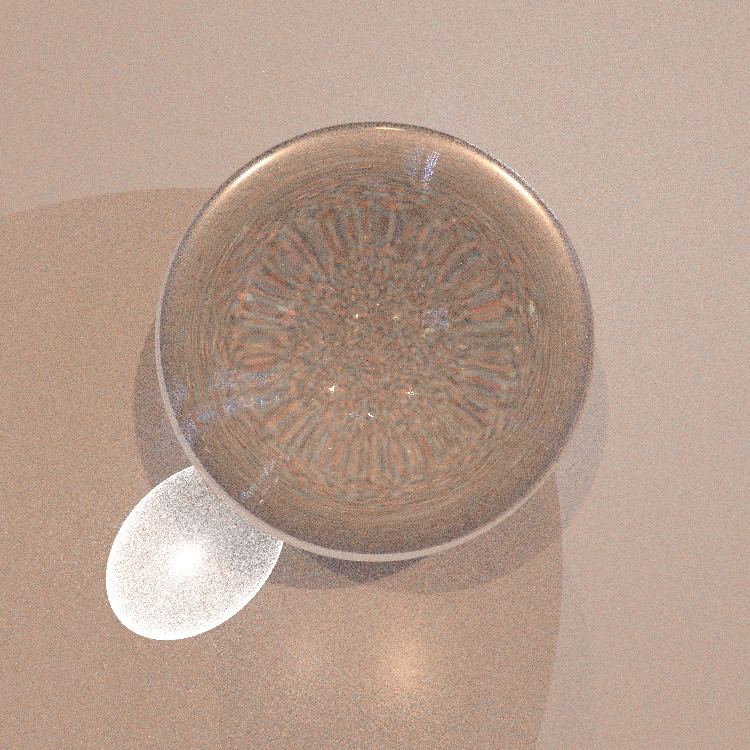} &
  \includegraphics[width=0.237\linewidth]{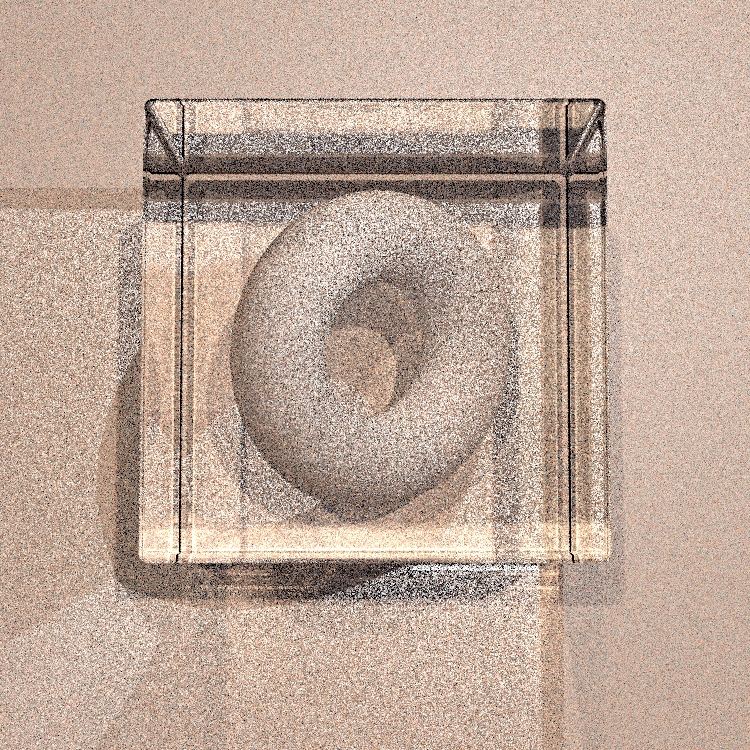} &
  \includegraphics[width=0.237\linewidth]{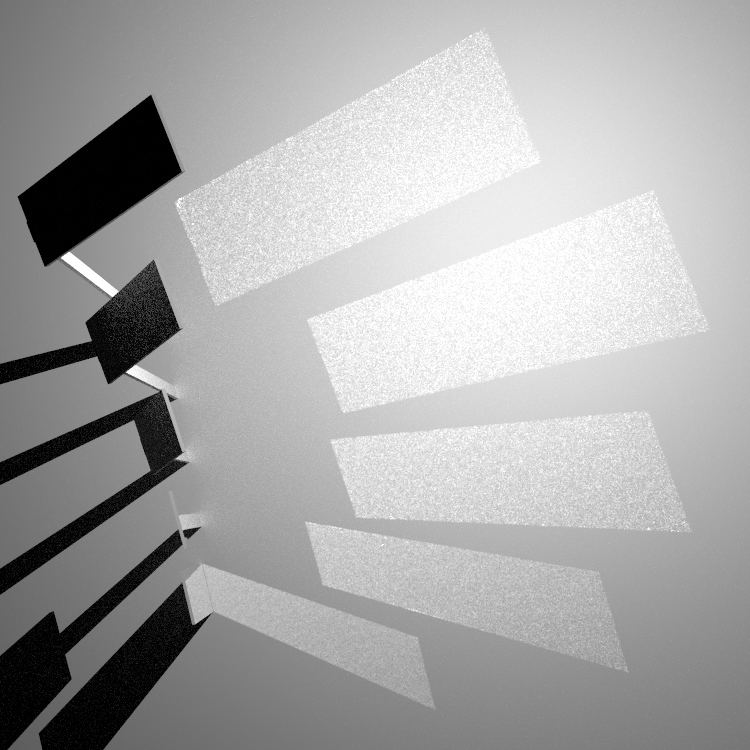} &
  \includegraphics[width=0.237\linewidth]{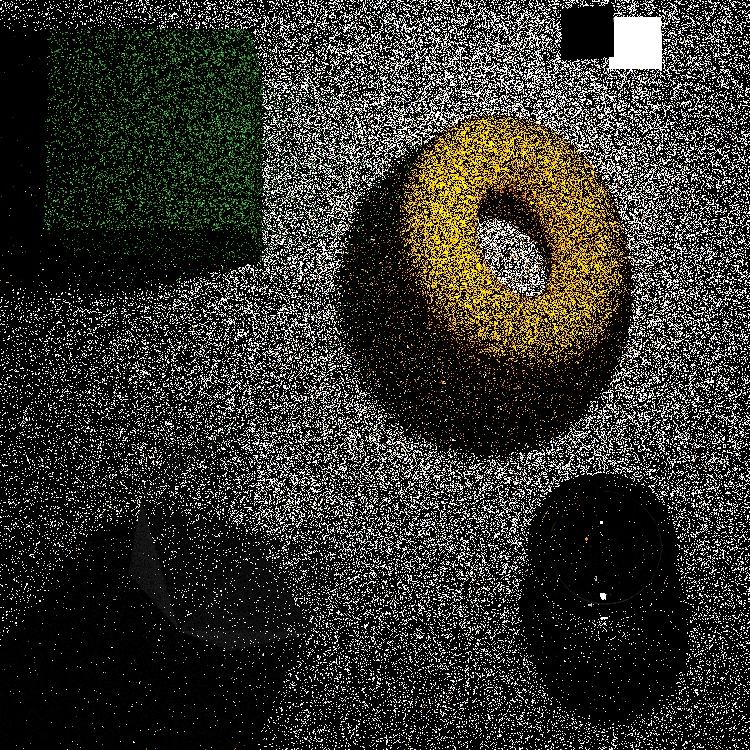}\\
\rotatebox{90}{\hspace{3.3em}PT} &
  \includegraphics[width=0.237\linewidth]{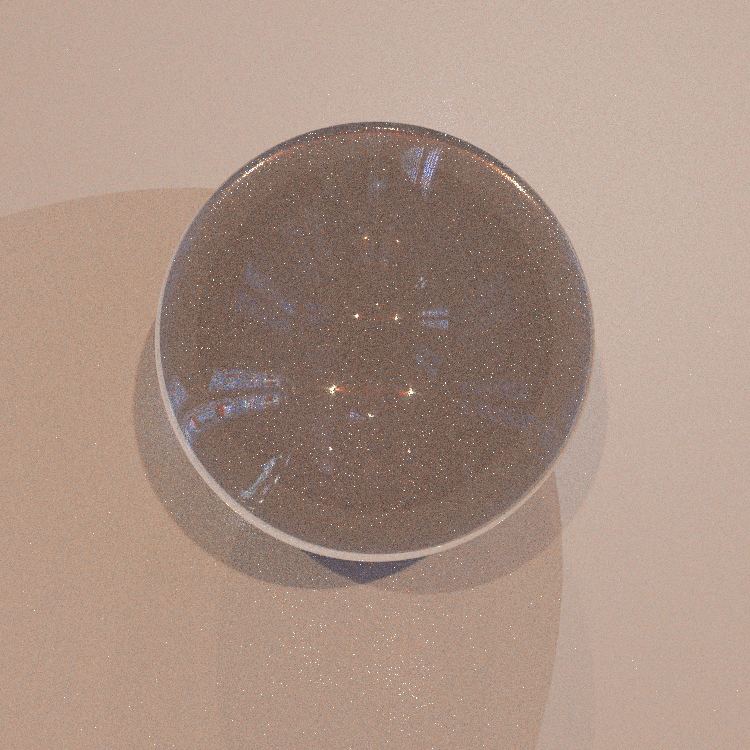} &
  \includegraphics[width=0.237\linewidth]{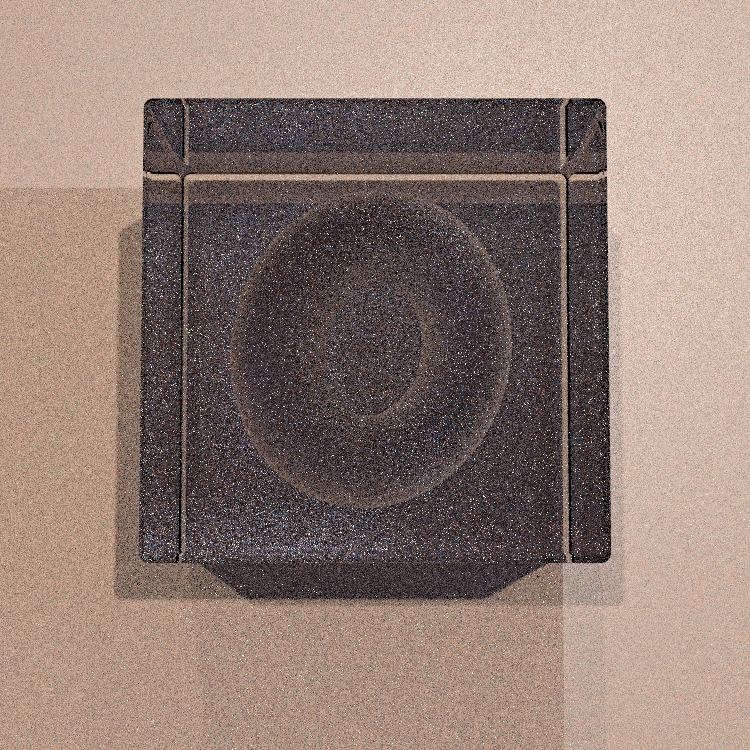} &
  \includegraphics[width=0.237\linewidth]{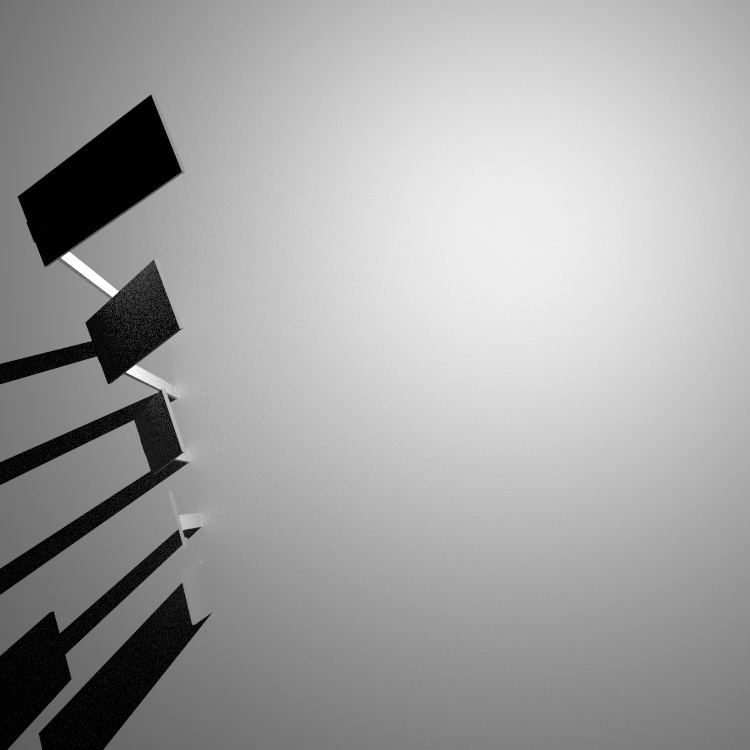} &
  \includegraphics[width=0.237\linewidth]{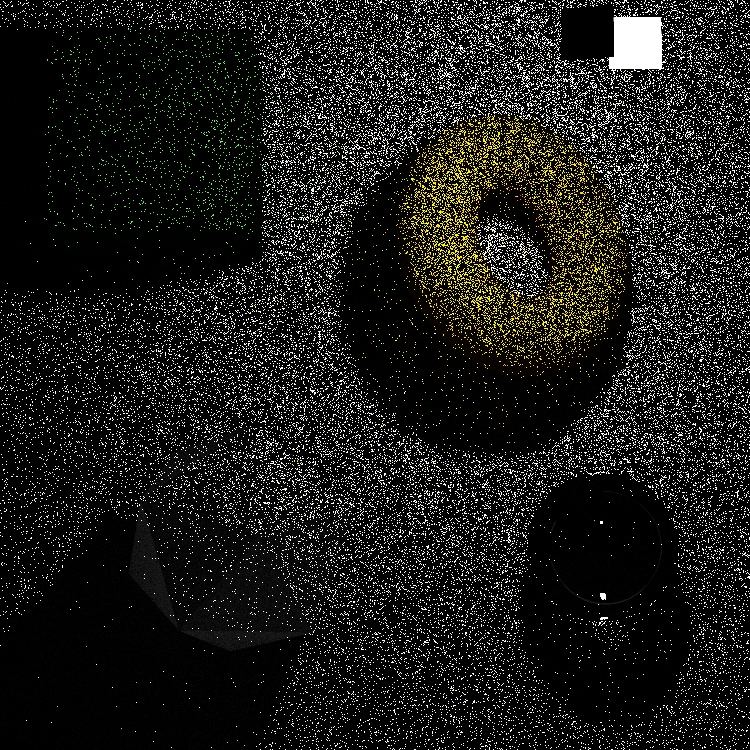}\\
\rotatebox{90}{\hspace{3.2em}BPT} &
  \includegraphics[width=0.237\linewidth]{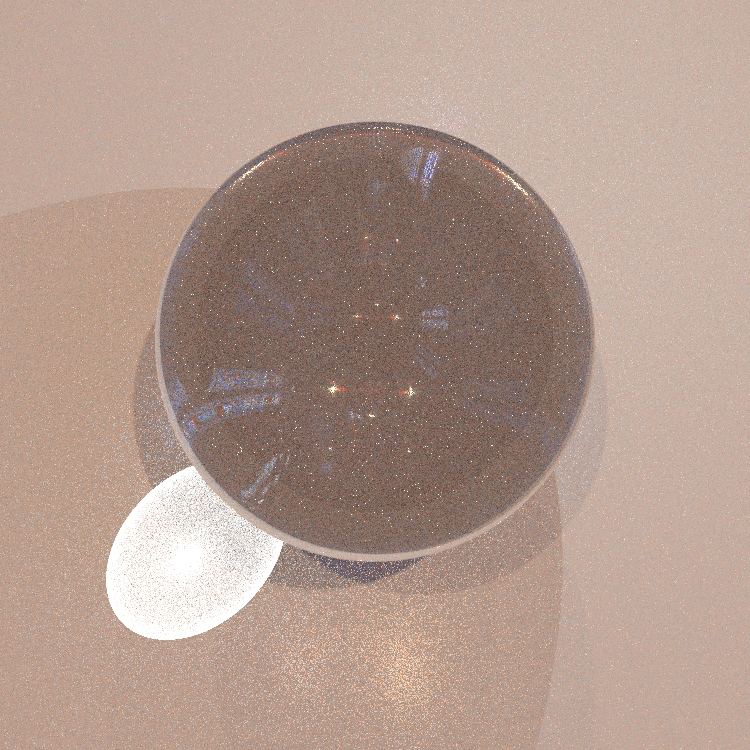} &
  \includegraphics[width=0.237\linewidth]{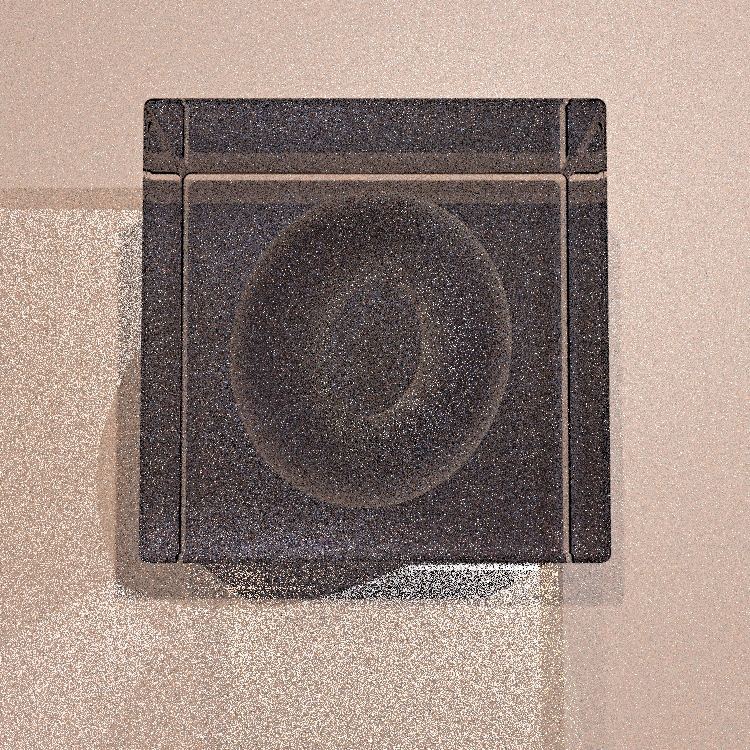} &
  \includegraphics[width=0.237\linewidth]{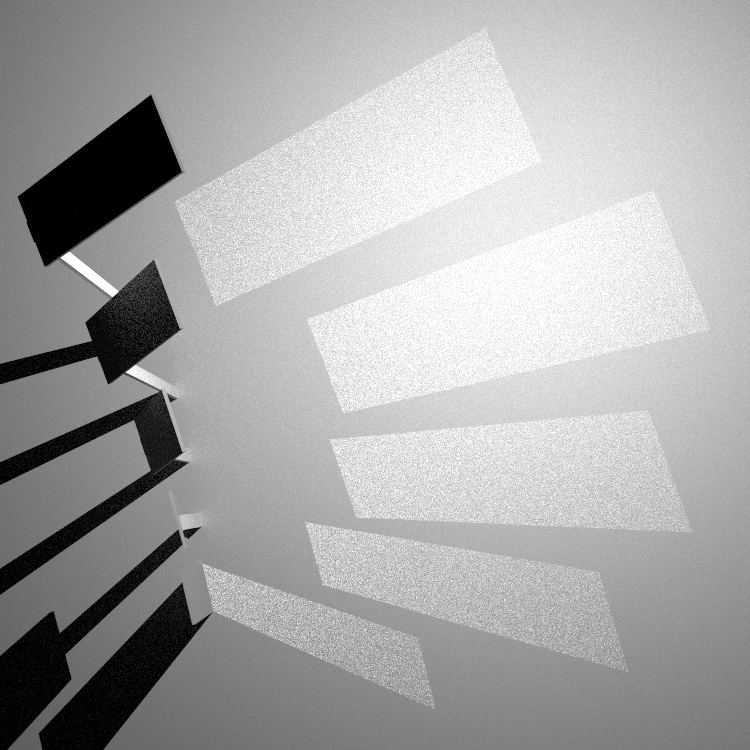} &
  \includegraphics[width=0.237\linewidth]{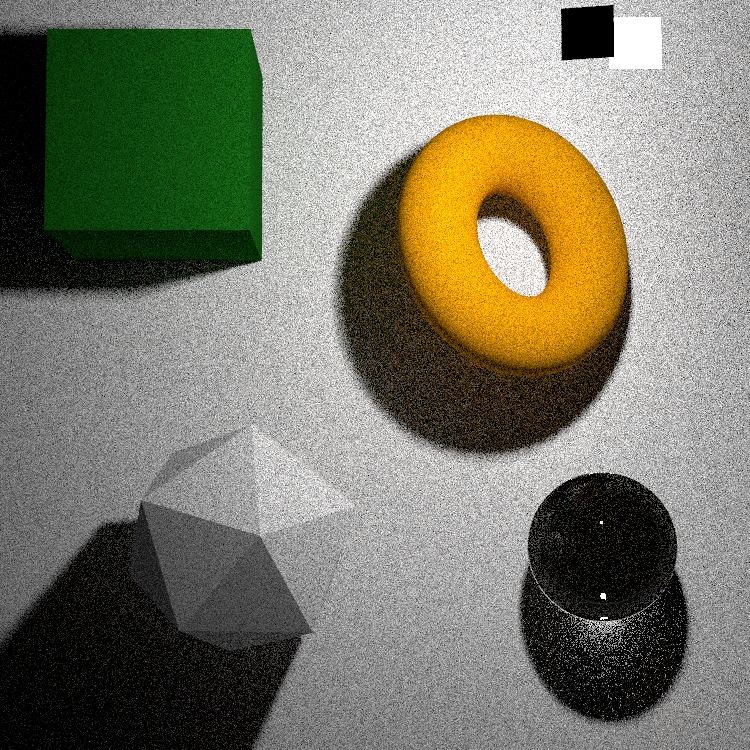}\\
\rotatebox{90}{\hspace{2.9em}BPM} &
  \includegraphics[width=0.237\linewidth]{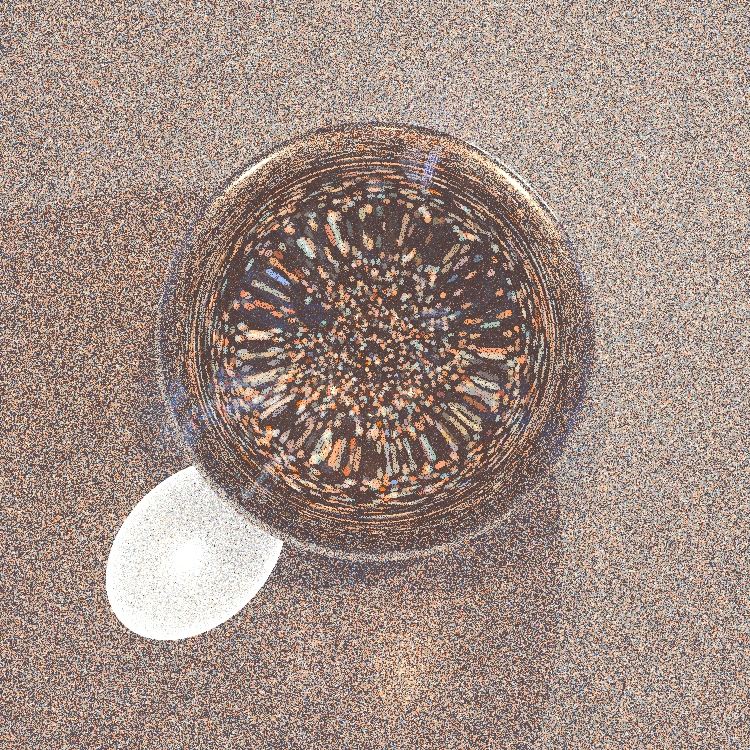} &
  \includegraphics[width=0.237\linewidth]{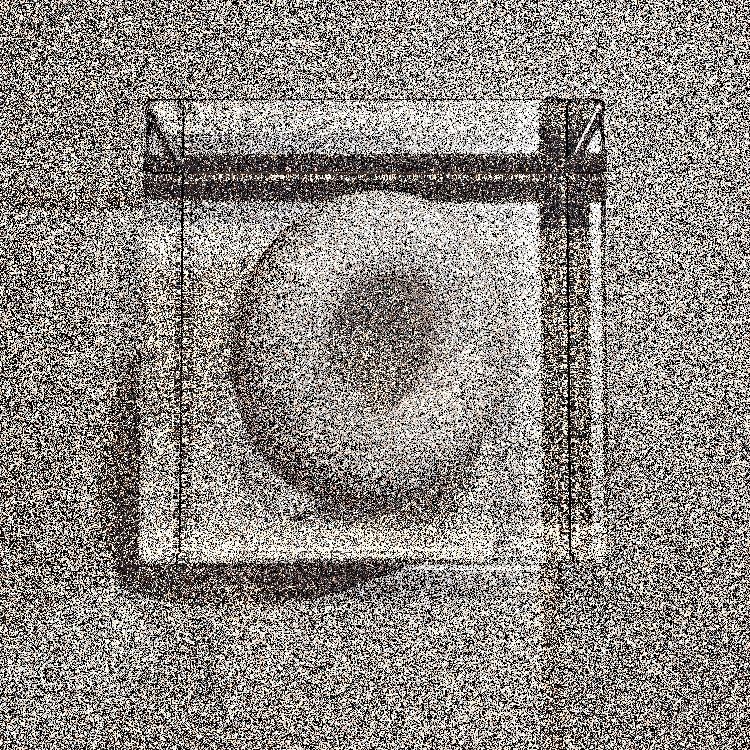} &
  \includegraphics[width=0.237\linewidth]{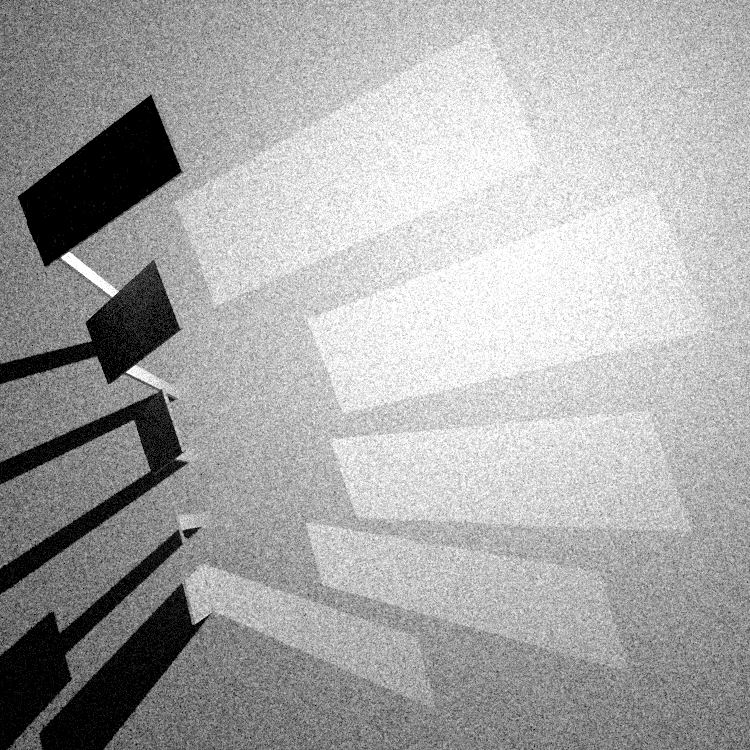} &
  \includegraphics[width=0.237\linewidth]{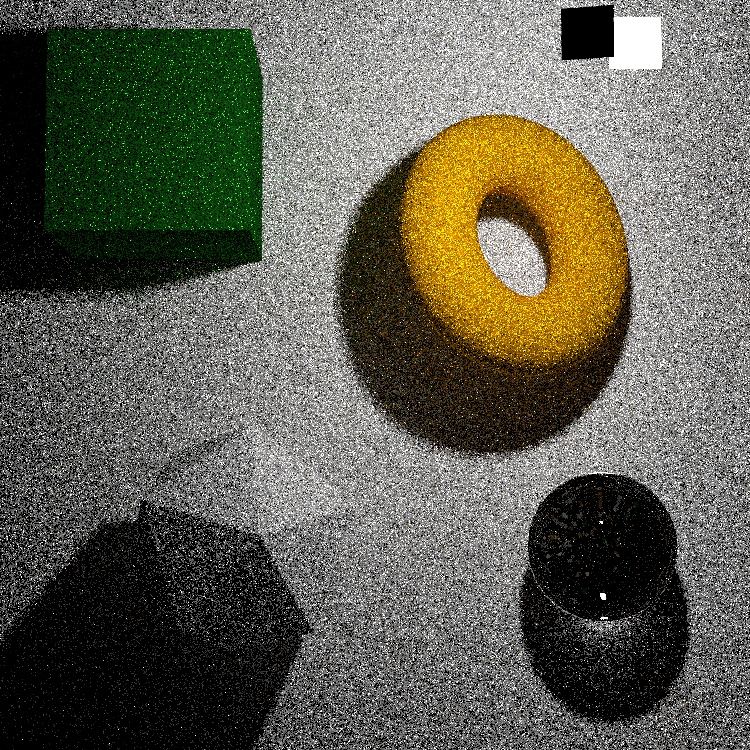}\\
\rotatebox{90}{\hspace{2.9em}VCM} &
  \includegraphics[width=0.237\linewidth]{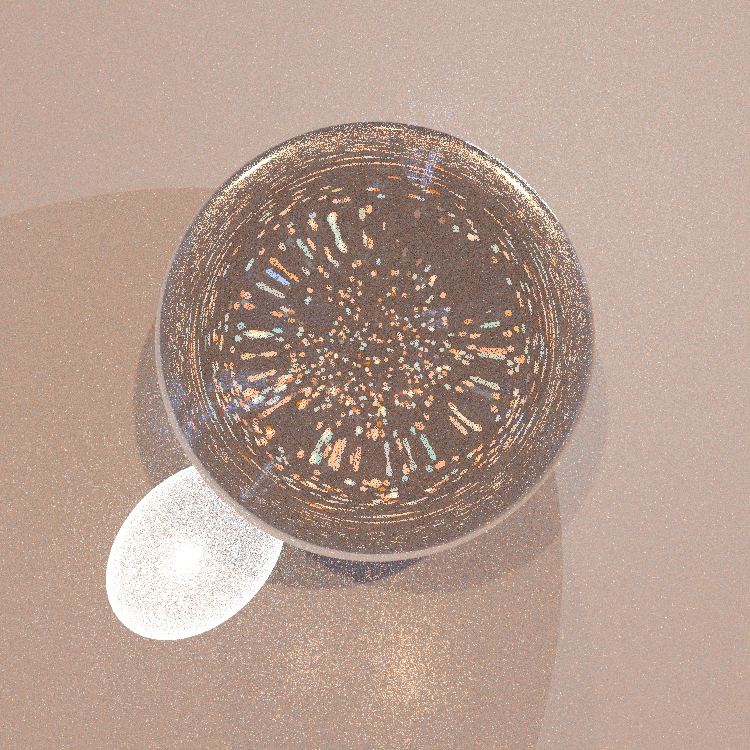} &
  \includegraphics[width=0.237\linewidth]{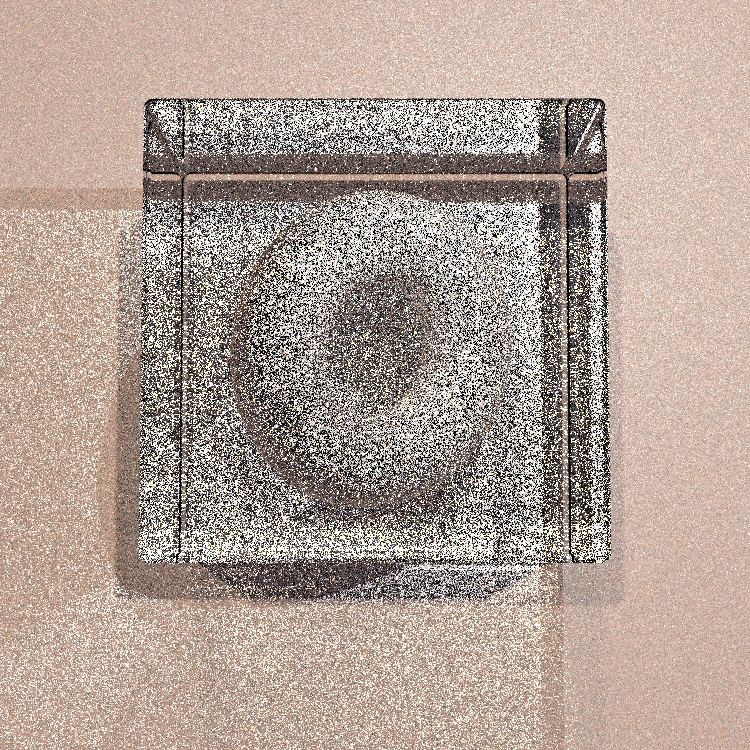} &
  \includegraphics[width=0.237\linewidth]{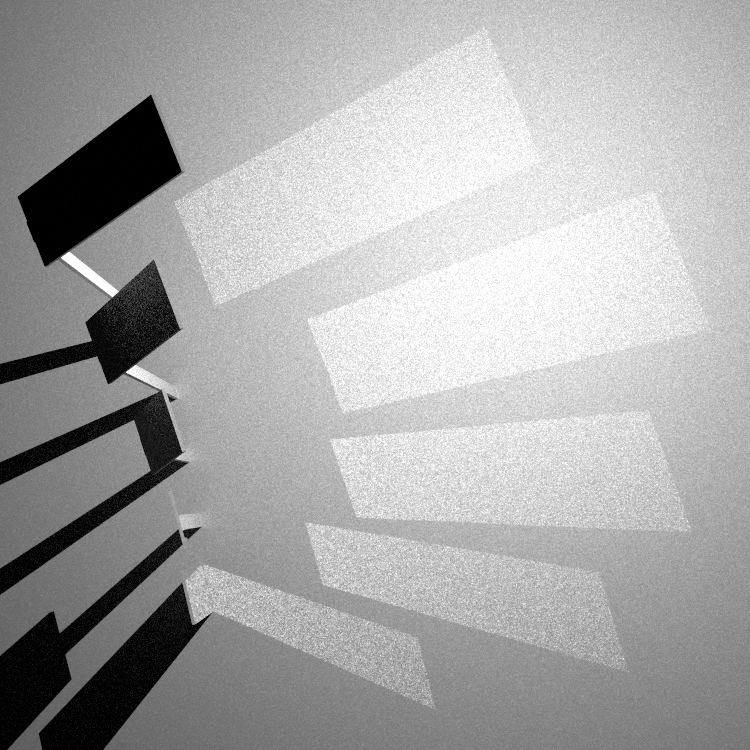} &
  \includegraphics[width=0.237\linewidth]{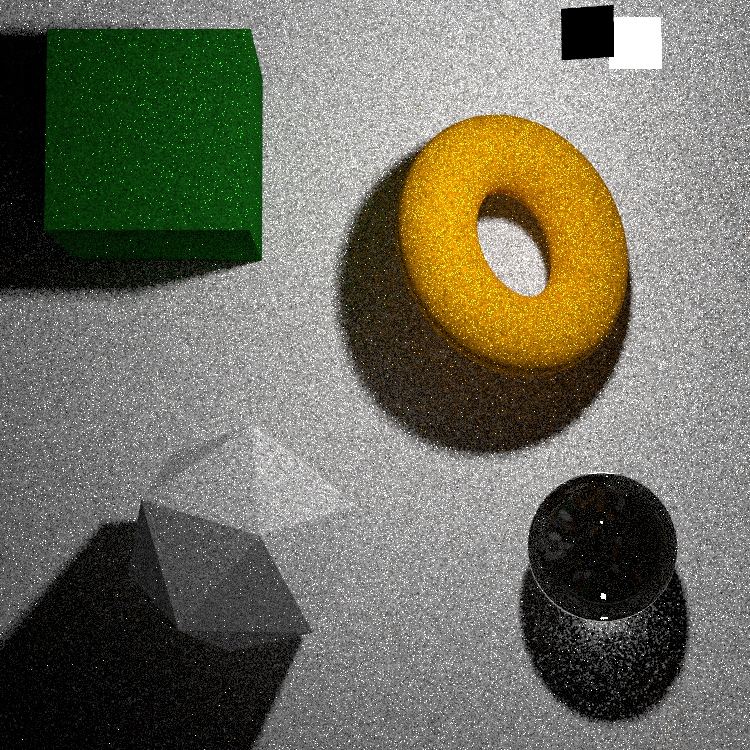}
\end{tabularx}

\caption{Equal time comparison (1 \si{\minute}) of different rendering algorithms in difficult light scenarios.}
\label{fig:sls}
 \end{figure}

In Figure \ref{fig:sls} we show selected light situations with an equal time comparison of the basic NEB method to other rendering algorithms.
For the first two scenarios our method is superior due to the uniform sampling density of NEE vertices on the glass surfaces.
The \textsc{Mirrors} scenario shows a small degeneration in quality where the distance to the caustic receiving surface increases (mirrors at the top).
The \textsc{Reflector} scenario represents the worst case.
Here, the tiny, bright surfaces close to the point light source are seldom found by a Path Tracer.

\begin{figure} \newcommand\spp[1]{\begin{tikzpicture}[overlay]%
	\node[anchor=south east,fill=black,text=white,inner sep=1pt,opacity=0.5,text opacity=1] at (0,0) {\scriptsize #1 spp};
\end{tikzpicture}}
\centering%
\setlength\tabcolsep{0.5pt}%
\begin{tabularx}{\linewidth}{Cccc}
& NEB & NEB + LP & VCM\\
\rotatebox{90}{\hspace{1.7em}\textsc{Watch} (\SI{5}{\minute})} &
	\includegraphics[width=0.316\linewidth]{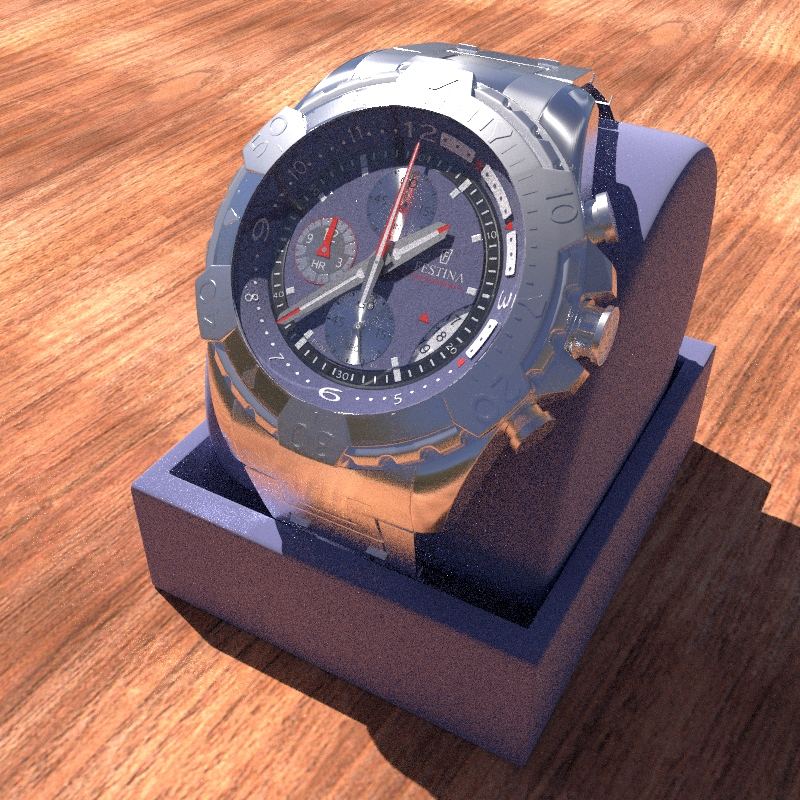}\spp{38} &
	\includegraphics[width=0.316\linewidth]{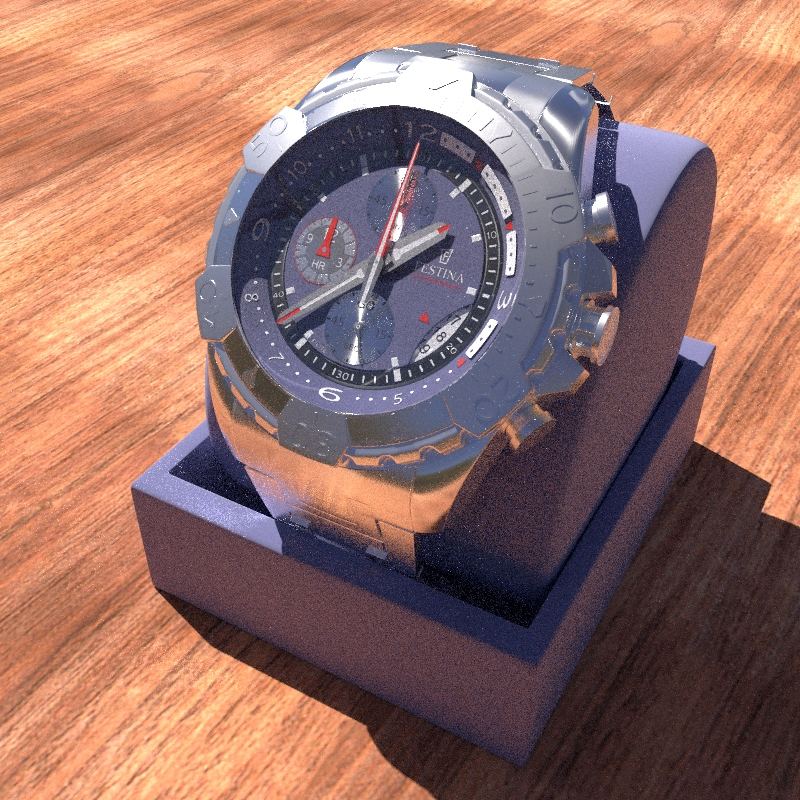}\spp{35} &
	\includegraphics[width=0.316\linewidth]{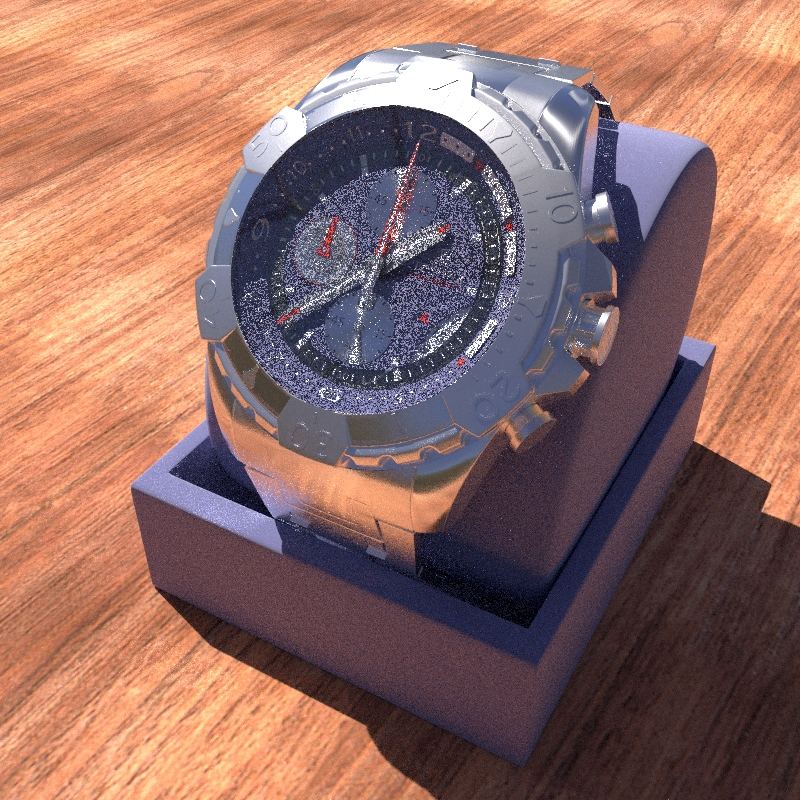}\spp{49}\\
	\rotatebox{90}{\hspace{0.6em}\textsc{Bathroom} (\SI{10}{\minute})} &
	\includegraphics[width=0.316\linewidth]{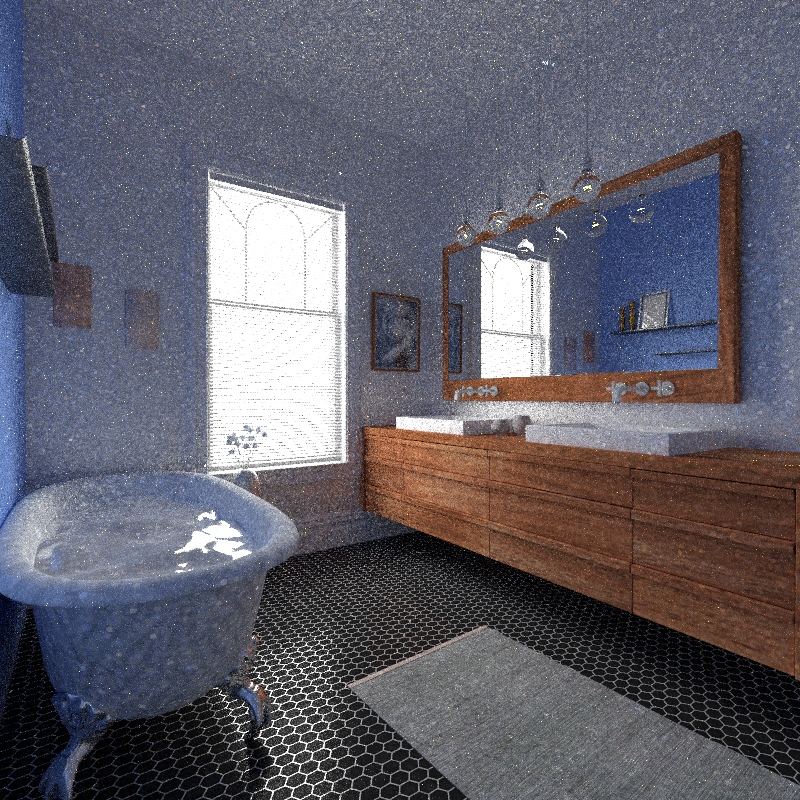}\spp{24} &
	\includegraphics[width=0.316\linewidth]{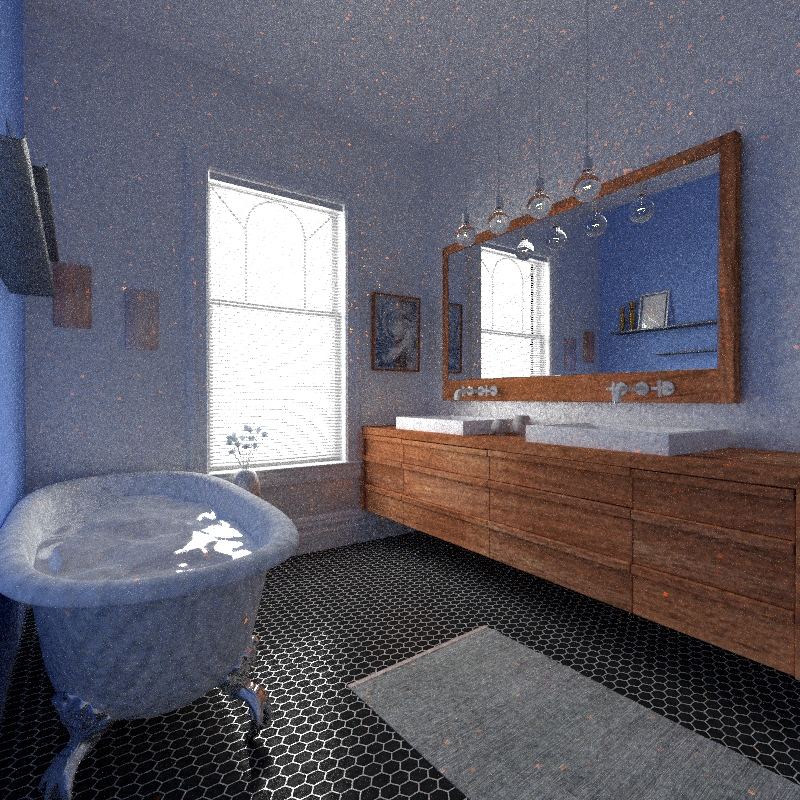}\spp{15} &
	\includegraphics[width=0.316\linewidth]{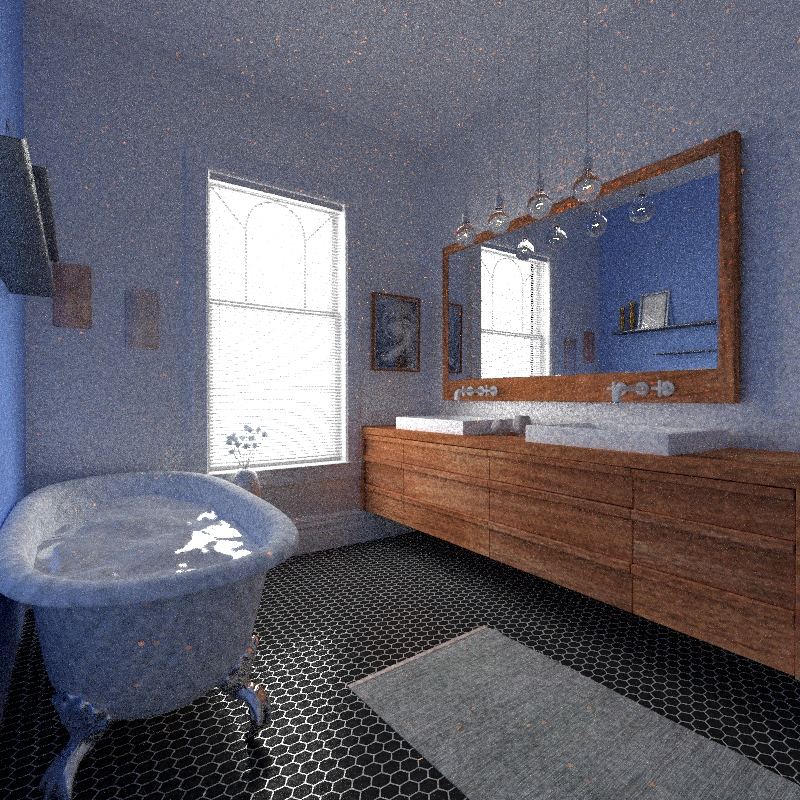}\spp{13}\\
	\rotatebox{90}{\hspace{0.6em}\textsc{Christmas} (\SI{15}{\minute})} &
	\includegraphics[width=0.316\linewidth]{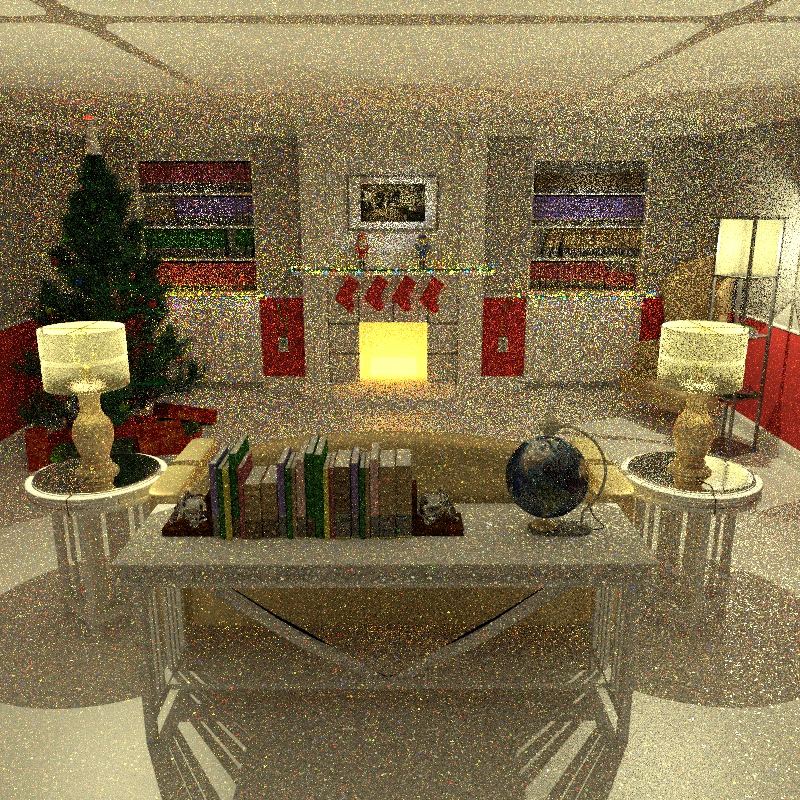}\spp{42} &
	\includegraphics[width=0.316\linewidth]{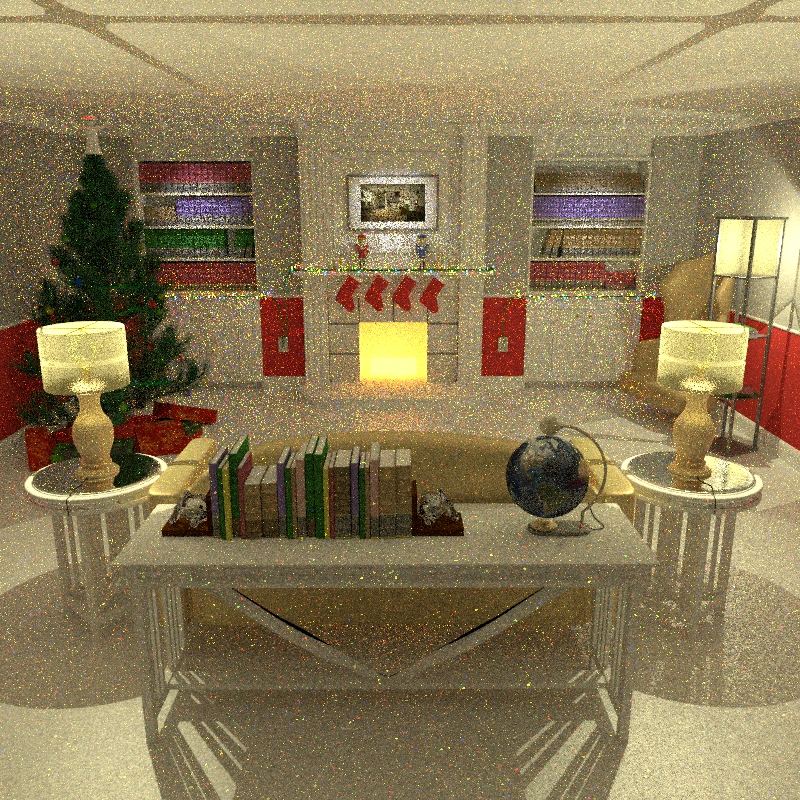}\spp{22} &
	\includegraphics[width=0.316\linewidth]{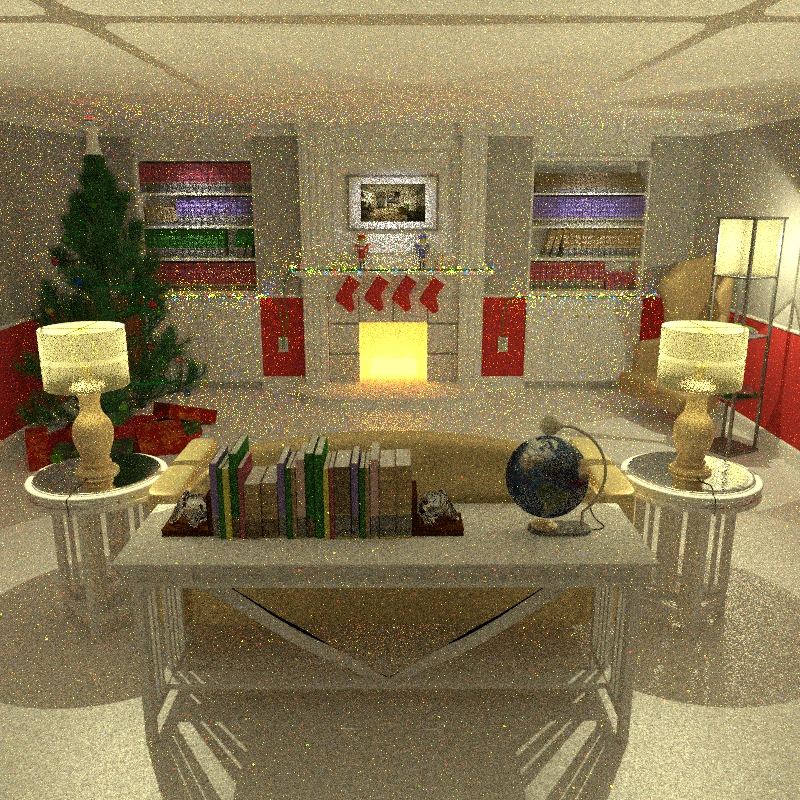}\spp{13}
\end{tabularx}

\caption{Comparison of NEB to VCM in more realistic scenes.}
\label{fig:realscenes}
 \end{figure}

In Figure \ref{fig:realscenes} more realistic scenes are compared.
For some situations like in the \textsc{Watch} scene, NEB is superior to the state of the art VCM.
In other situations (\textsc{Bathroom} or \textsc{Christmas} scene) it shows more noise than VCM without modifications.
Enabling the additional tracing of light photons makes NEB equally effective as VCM.
In all our experiments we found that NEB+LP is very strong for caustics and SDS paths regardless of the scene scale.
For diffuse indirect lighting it performs on an average level which is comparable to most other methods.

Comparing NEB as a method to produce high quality photons maps against the older methods \cite{peter_importance_1998,suykens_density_2000,keller_efficient_2000}, it produces a medium quality map.
In most scenes, it performs similarly to the other methods without wasting samples for the estimation of importance distributions.
However, if there are small reflectors close to the light source (light bulb or Figure \ref{fig:sls}) it will be less effective, because it does not respect the radiance distribution.
Adding standard photons compensates this weakness, but their distribution does not follow importance again.
However, since none of the photons is ever stored, the memory consumption of additional photons is of no interest.

\section{Conclusions and Future Work}
\label{sec:conclusions}

We proposed a new light transport operator -- the next event backtracking -- which is strong in situations where the usual photon transport fails.
It can successfully render caustics of small objects in large scenes due to an implicit guidance of light paths toward the important regions and is often very strong for SDS paths.

The most difficult problem is a robust and fast density estimate to convert irradiance into flux at the selected positions.
We introduced an octree-based data structure which is fast, but may cause small artifacts from its grid structure leading to overly bright photons.

Our NEB algorithm used the operator in the setup of a conventional Path Tracer.
In this configuration, certain light situations can still cause high variance.
The modification to add conventional photon tracing, beginning at the light source, solves this problem.
Effectively, NEB is weak if standard photon tracing is strong and vice versa.
Therefore, both together result in a very robust algorithm.
Only scenes with bad visibility between observer and light source are still a problem.

One of the practical weaknesses is the memory consumption, which is slightly higher than in other photon mapping-based approaches.
The view path vertex storage exchanges the one for photons and has a comparable size.
Additionally, there is the octree to store the density values with up to 50 MB and the storage for emissive events with another 50 MB per one million paths.

Also, it would be interesting to implement the algorithm on a GPU.
In contrast to BPT or VCM, the number of connections is only linear in path lengths.
This reduces divergence and per-path-time such that our method should scale well on parallel hardware.

Another future avenue would be to use MCMC samplers for the view paths or to use the NEB paths as seed paths for MCMC samplers.
This would increase the method's robustness with respect to bad visibility of the light source due to high occlusion.

Finally, our approach allows the use of arbitrarily sampled positions as photon emitters.
It would be possible (but likely ineffective) to distribute emitters equally on the surfaces -- independent of camera and light sources.
More interesting would be to use Markov chains to sample the emitters on the surfaces themselves.
Similar to Lightweight Photonmapping \cite{grittmann_efficient_2018}, it would then be possible to remove emitters on surfaces where other samplers are more successful and to increase the density otherwise.

\section*{Acknowledgments}

I want to thank those people who make 3D scenes publicly available.
The villa and the bathroom scene are taken form the PBRT repository.
The stadium was modeled by Ericchip1983 and is provided on \href{	https://www.blendswap.com/blends/74526}{blendswap.com/blends/74526}.
The wrist watch was created by HeraSK (\href{https://www.blendswap.com/blends/70232}{blendswap.com/blends/70232}).


\bibliographystyle{alpha}
{\footnotesize
\bibliography{nexteventbacktracking}}

\pagebreak
\appendix
\section{Intersection Area between Plane and Box}
\label{app:area}

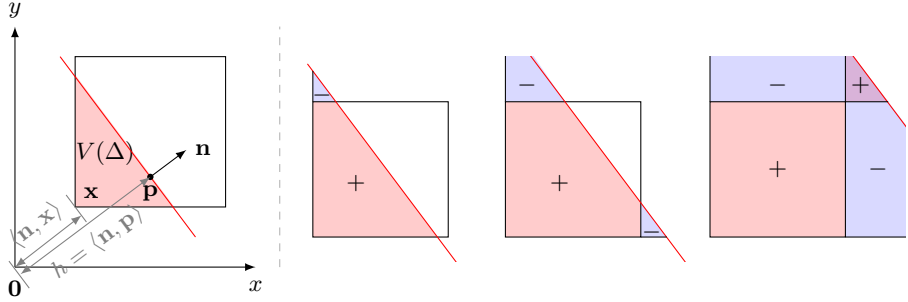
\begin{figure}
	\begin{tikzpicture}[scale=0.8]
	\small
		\begin{scope}
			\draw[->] (0,0) -- (4,0) node[anchor=north,yshift=-2pt] {$x$};
			\draw[->] (0,0) -- (0,4) node[anchor=south] {$y$};
			\node[anchor=north,yshift=-2pt] at (0,0) {$\mvec{0}$};
			
			\fill[red!20] (1,1) -- (2.65,1) -- (1,3.2) -- cycle;
			\node at (1.5,1.9) {$V(\Delta)$};
			\draw (1,1) rectangle +(2.5,2.5);
			\node[anchor=south west] at (1,1) {$\mvec{x}$};
			\draw[red] (0.75, 3.5) -- (3,0.5);
			\fill (2.25,1.5) circle [radius=1.5pt] node[anchor=north] {$\mvec{p}$};
			\draw[->] (2.25,1.5) -- +(3*0.2,2.25*0.2) node[anchor=west] {$\mvec{n}$};
			\draw[black!50] (0.85, 1.2) -- +(2.25*0.15,-3*0.15);
			\draw[<->,black!50] (0,0) -- +(3*0.37,2.27*0.38) node[pos=0.5,above,sloped] {$\langle\mvec{n},\mvec{x}\rangle$};
			\draw[black!50] (-0.1, 0.1) -- +(2.25*0.15,-3*0.15);
			\draw[<->,black!50] (2.25,1.5) -- +(-3*0.72,-2.25*0.72) node[pos=0.5,below,sloped] {$h=\langle\mvec{n},\mvec{p}\rangle$};
		\end{scope}
		\draw[black!30,dashed] (4.4,0) -- (4.4,4);
		\clip (4.3,0.1) rectangle +(10.5,3.4);
		\begin{scope}[xshift=4.5cm,yshift=-0.4cm,scale=0.9]
			\fill[red!20] (0.5,1) -- (2.8,1) -- (0.92,3.5) -- (0.5,3.5);
			\fill[blue!15] (0.5,3.5) -- (0.92,3.5) -- (0.5,4.07);
			\draw (0.5,1) rectangle +(2.5,2.5);
			\draw[red] (0.4, 4.2) -- +(2.25*1.3,-3*1.3);
			\draw (0.5,3.5) -- (0.5,4.07);
			\node at (1.3,2) {$+$};
			\node at (0.67,3.63) {$-$};
		\end{scope}
		\begin{scope}[xshift=7.7cm,yshift=-0.4cm,scale=0.9]
			\fill[red!20] (0.5,1) -- (3,1) -- (3,1.63) -- (1.6,3.5) -- (0.5,3.5);
			\fill[blue!15] (0.5,3.5) -- (1.6,3.5) -- (0.5,5.1);
			\fill[blue!15] (3,1) -- (3,1.63) -- (3.48,1);
			\draw (0.5,1) rectangle +(2.5,2.5);
			\draw[red] (0.4, 5.1) -- +(2.25*1.6,-3*1.6);
			\draw (0.5,3.5) -- (0.5,5.0) (3,1) -- (3.48,1);
			\node at (1.5,2) {$+$};
			\node at (0.9,3.8) {$-$};
			\node at (3.2,1.1) {$-$};
		\end{scope}
		\begin{scope}[xshift=11.1cm,yshift=-0.4cm,scale=0.9]
			\fill[red!20] (0.5,1) rectangle (3,3.5);
			\fill[blue!15] (0.5,3.5) -- (3,3.5) -- (3,4.54) -- (0.5,8);
			\fill[blue!15] (3,1) -- (5.7,1) -- (3.79,3.5) -- (3,3.5);
			\fill[red!50!blue!30] (3,3.5) -- (3.79,3.5) -- (3,4.54);
			\draw (0.5,1) rectangle +(2.5,2.5);
			\draw[red] (0.4, 8.0) -- +(2.25*5,-3*5);
			\draw (0.5,3.5) -- (0.5,5.0) (3,1) -- (5.0,1)
				(3,3.5) -- (3.79,3.5) (3,3.5) -- (3,4.54);
			\node at (1.75,2.25) {$+$};
			\node at (1.75,3.8) {$-$};
			\node at (3.6,2.25) {$-$};
			\node at (3.28,3.8) {$+$};
		\end{scope}
	\end{tikzpicture}
	\caption{Intersection area from simplex volume.
	In 2D the simplex-volume is the area of a triangle.
	Left: a single vertex of the box lies below the plane ($\mvec{p},\mvec{n}$).
	Right: sequence of events when moving the plane along $\mvec{n}$.}
	\label{fig:simplex}
\end{figure}

We are sure that this formula is not new, but we were not able to find a cite-able reference for Equation \eqref{eq:area}.
We found the derivation on Mathoverflow ({\footnotesize\url{https://math.stackexchange.com/a/885662/661978}}) and repeat it here for completeness.

The basic idea is to compute the volume $V$ inside the box and below the plane with respect to the plane offset $h = \langle \mvec{n},\mvec{p} \rangle$.
Then the derivation of $V$ yields the searched area.
As primary conditions we have $\lVert\mvec{n}\rVert = 1$ and that the box is axis aligned.
Let
\begin{equation}
	\Delta(h, \mvec{x}) = \{ \mvec{y} \in \mathbb{R}^d : \forall \mvec{y}_i \geq \mvec{x}_i \wedge \langle\mvec{n},\mvec{y}\rangle\leq h \}
\end{equation}
be the $d$-dimensional simplex, which starts at $\mvec{x}$ as depicted in Figure \ref{fig:simplex}.
The volume of this simplex can be computed from a product of its (axis aligned) edge lengths with
\begin{equation}
	V(\Delta(h, \mvec{x})) = V(\Delta(h-\langle\mvec{n},\mvec{x}\rangle, \mvec{0})) = \frac{\max(0,h - \langle\mvec{n},\mvec{x}\rangle)^d}{d!\prod_i \mvec{n}_i}.
\end{equation}
The clamping to zero is necessary if the entire box is on the upper side of the plane.

Now, moving the plane along $\mvec{n}$, it will pass all vertices of the box in a sorted order.
This sequence is shown in Figure \ref{fig:simplex}, too.
After passing the second vertex, the simplex from the first one will overestimate the volume.
Here, the second simplex starting at the second vertex must be subtracted.
The same applies to the third vertex.
After the forth vertex, the subtracted areas from the second and third vertex will overlap and the volume is underestimated.
Thus, the volume of the forth simplex must be added again.
This alternating sign is described by the parity $\epsilon$ as given in Equation \eqref{eq:area}.

Hence, the volume of the box $\mathcal{B}$ with respect to the plane is
\begin{equation}
	V(\mathcal{B},h) = \frac{1}{d!\prod_i \mvec{n}_i} \sum\limits_{i=0}^{2^d-1} \epsilon_i \max(0,h - \langle\mvec{n},\mvec{b}_i\rangle)^d.
\end{equation}
Finally, the change of volume over $h$ is dominated by the area of the infinitesimal slab which gives the area
\begin{align}
	A(\mathcal{B},h) &= \frac{\partial V(\mathcal{B},h)}{\partial h}\nonumber \\
	&= \frac{1}{(d-1)!\prod_i \mvec{n}_i} \sum\limits_{i=0}^{2^d-1} \epsilon_i \max(0,h - \langle\mvec{n},\mvec{b}_i\rangle)^{(d-1)}.
\end{align}

Equation \eqref{eq:area} is then obtained by applying the above equation for one, two and three dimensions.
Each component of $\mvec{n}$ which is zero means that the normal is perpendicular to the respective edge of the box.
Then, the size $\mvec{s}$ of the box must be multiplied with the area of the simplex from the reduced dimension.

\pagebreak
\section{Octree Implementation}
\label{app:octreeimpl}

To use the following C++ source codes you must provide some implementation of a 3D vector with following operations:
\begin{itemize}
	\setlength\itemsep{0pt}
	\item arithmetic (+, -, *, /)
	\item array access [0] to [2] for the three dimensions
	\item functions like dot(), abs(), len()
\end{itemize}

\noindent Helper function which implements Equation \eqref{eq:area}:
\begin{lstlisting}
inline float sq(float x) { return x * x; }

// Compute the area of the plane-box intersection
// https://math.stackexchange.com/questions/885546
// https://math.stackexchange.com/a/885662
// or appendix A
inline float intersection_area(const vec3& bmin, const vec3& bmax,
															 const vec3& pos, const vec3& normal) {
	vec3 cellSize = bmax - bmin;
	vec3 absN = abs(normal);
	// 1D cases
	if(abs(absN[0] - 1.0f) < 1e-3f) return cellSize[1] * cellSize[2];
	if(abs(absN[1] - 1.0f) < 1e-3f) return cellSize[0] * cellSize[2];
	if(abs(absN[2] - 1.0f) < 1e-3f) return cellSize[0] * cellSize[1];
	// 2D cases
	for(int d = 0; d < 3; ++d) if(absN[d] < 1e-4f) {
		int x = (d + 1) % 3;
		int y = (d + 2) % 3;
		// Use the formula from stackexchange: phi(t) = max(0,t)^2 / 2 m_1 m_2
		// -> l(t) = sum^4 s max(0,t-dot(m,v)) / m_1 m_2
		// -> A(t) = l(t) * h_3
		float t = normal[x] * pos[x] + normal[y] * pos[y];
		float sum = 0.0f;
		sum += max(0.0f, t - (normal[x] * bmin[x] + normal[y] * bmin[y]));
		sum -= max(0.0f, t - (normal[x] * bmin[x] + normal[y] * bmax[y]));
		sum -= max(0.0f, t - (normal[x] * bmax[x] + normal[y] * bmin[y]));
		sum += max(0.0f, t - (normal[x] * bmax[x] + normal[y] * bmax[y]));
		return cellSize[d] * abs(sum / (normal[x] * normal[y]));
	}
	// 3D cases
	float t = dot(normal, pos);
	float sum = 0.0f;
	sum += sq(max(0.0f, t - dot(normal, bmin)));
	sum -= sq(max(0.0f, t - dot(normal, vec3{bmin[0], bmin[1], bmax[2]})));
	sum += sq(max(0.0f, t - dot(normal, vec3{bmin[0], bmax[1], bmax[2]})));
	sum -= sq(max(0.0f, t - dot(normal, vec3{bmin[0], bmax[1], bmin[2]})));
	sum += sq(max(0.0f, t - dot(normal, vec3{bmax[0], bmax[1], bmin[2]})));
	sum -= sq(max(0.0f, t - dot(normal, vec3{bmax[0], bmin[1], bmin[2]})));
	sum += sq(max(0.0f, t - dot(normal, vec3{bmax[0], bmin[1], bmax[2]})));
	sum -= sq(max(0.0f, t - dot(normal, bmax)));
	return abs(sum / (2.0f * normal[0] * normal[1] * normal[2]));
}
\end{lstlisting}

\noindent Helper to track the depth of the tree:
\begin{lstlisting}
template<typename T>
inline void atomic_max(std::atomic<T>& a, T b) {
	T oldV = a.load();
	while(oldV < b && !a.compare_exchange_weak(oldV, b)) ;
}
\end{lstlisting}

\begin{lstlisting}
// A sparse octree with atomic insertion to measure the density of elements
// in 3D space.
class DensityOctree {
	static constexpr int SPLIT_FACTOR = 4;
	// At some time the counting should stop -- otherwise the counter will
	// overflow inevitable.
	static constexpr int FILL_ITERATIONS = 1000;
public:
	void set_iteration(int iter) {
		int iterClamp = min(FILL_ITERATIONS, iter);
		m_stopFilling = iter > FILL_ITERATIONS;
		m_densityScale = 1.0f / iterClamp;
		m_splitCountDensity = SPLIT_FACTOR * iterClamp;
		// Set the counter of all unused cells to the number of expected samples
		// divided by 4. A planar surface will never extend to all eight cells.
		// It might intersect 7 of them, but still the distribution is on a
		// surface. Therefore, the SPLIT_FACTOR many particles are distribute
		// among 4 cells. This gives a much better value than dividing the
		// factor by 8.
		if(!m_stopFilling)
			for(int i = m_allocationCounter.load(); i < m_capacity; ++i)
				m_nodes[i].store(ceil(SPLIT_FACTOR / 4.0f * iter));
	}

	void initialize(const vec3& sceneMin, const vec3& sceneMax, int capacity) {
		// Slightly enlarge the volume to avoid numerical issues on the boundary
		vec3 sceneSize = (sceneMax - sceneMin) * 1.002f;
		m_sceneSizeInv = 1.0f / sceneSize;
		m_sceneScale = len(sceneSize);
		m_minBound = sceneMin - sceneSize * (0.001f / 1.002f);
		// Round up to 8n+1 - otherwise we cannot use the last [1,7] entries.
		m_capacity = 1 + ((capacity + 7) & (~7));
		m_nodes = std::make_unique<std::atomic_int32_t[]>(m_capacity);;
		// Allocate the root node with a count of 0
		m_allocationCounter.store(1);
		m_nodes[0].store(0);
		m_depth.store(0);
	}

	// Overwrite all counters with 0, but keep allocation and child pointers.
	void clear_counters() {
		int n = m_allocationCounter.load();
		for(int i = 0; i < n; ++i)
			if(m_nodes[i].load() > 0)
				m_nodes[i].store(0);
	}

	void increment(const vec3& pos) {
		if(m_stopFilling) return;
		vec3 normPos = (pos - m_minBound) * m_sceneSizeInv;
		int countOrChild = increment_if_positive(0);
		countOrChild = split_node_if_necessary(0, countOrChild, 0);
		int edgeL = 1;
		int currentDepth = 0;
		while(countOrChild < 0) {
			edgeL *= 2;
			++currentDepth;
			// Get the relative index of the child [0,7]
			ivec3 intPos = (ivec3{ normPos * edgeL }) & 1;
			int idx = intPos[0] + 2 * (intPos[1] + 2 * intPos[2]);
			idx -= countOrChild;	// 'Add' global offset (which is stored negative)
			countOrChild = increment_if_positive(idx);
			countOrChild = split_node_if_necessary(idx, countOrChild, currentDepth);
		}
	}






	float get_density(const vec3& pos, const vec3& normal, float* size = 0) {
		vec3 offPos = pos - m_minBound;
		vec3 normPos = offPos * m_sceneSizeInv;
		// Get the integer position on the finest level.
		int gridRes = 1 << m_depth.load();
		ivec3 iPos { normPos * gridRes };
		// Get root value. This will most certainly be a child pointer...
		int countOrChild = m_nodes[0].load();
		// The most significant bit in iPos distinguishes the children of the
		// root node. For each level, the next bit will be the relevant one.
		int currentLvlMask = gridRes;
		while(countOrChild < 0) {
			currentLvlMask >>= 1;
			// Get the relative index of the child [0,7]
			int idx = ((iPos[0] & currentLvlMask) ? 1 : 0)
					+ ((iPos[1] & currentLvlMask) ? 2 : 0)
					+ ((iPos[2] & currentLvlMask) ? 4 : 0);
			// 'Add' global offset (which is stored negative)
			idx -= countOrChild;
			countOrChild = m_nodes[idx].load();
		}
		if(countOrChild > 0) {
			// Get the world space cell boundaries
			int currentGridRes = gridRes / currentLvlMask;
			ivec3 cellPos = iPos / currentLvlMask;
			vec3 cellSize = 1.0f / (currentGridRes * m_sceneSizeInv);
			vec3 cellMin = cellPos * cellSize;
			vec3 cellMax = cellMin + cellSize;
			float area = intersection_area(cellMin, cellMax, offPos, normal);
			// Sometimes the above method returns zero. Therefore, we restrict the
			// area to something larger than 1/100 of an approximate cell area.
			float cellDiag = m_sceneScale / currentGridRes;
			float minArea = cellDiag * cellDiag;
			if(size) { *size = cellDiag; minArea *= 0.1f; }
			else minArea *= 0.01f;
			return m_densityScale * countOrChild / max(minArea, area);
		}
		return 0.0f;
	}

	// The robust version additional samples four neighbors in the tangent
	// plane and returns the median of those results. This removes noise and
	// therefore outliers in the rendering.
	float get_density_robust(const vec3& pos, const scene::TangentSpace& ts) {
		float d[5];
		float cellDiag = 1e-3f;
		int count = 0;
		d[0] = get_density(pos, ts.geoN, &cellDiag);
		cellDiag *= 1.1f;
		if(d[0] > 0.0f) ++count;
		d[count] = get_density(pos + ts.shadingTX * cellDiag, ts.geoN);
		if(d[count] > 0.0f) ++count;
		d[count] = get_density(pos - ts.shadingTX * cellDiag, ts.geoN);
		if(d[count] > 0.0f) ++count;
		d[count] = get_density(pos + ts.shadingTY * cellDiag, ts.geoN);
		if(d[count] > 0.0f) ++count;
		d[count] = get_density(pos - ts.shadingTY * cellDiag, ts.geoN);
		if(d[count] > 0.0f) ++count;
		// Find the median via selection sort up to the element m.
		// Prefer the greater element, because overestimations do not
		// cause such visible artifacts.
		int m = count / 2;
		for(int i = 0; i <= m; ++i) for(int j = i+1; j < count; ++j)
			if(d[j] < d[i])
				std::swap(d[i], d[j]);
		return d[m];
	}

	int capacity() const { return m_capacity; }
	int size() const { return min(m_capacity, m_allocationCounter.load()); }


private:
	// Nodes consist of 8 atomic counters OR child indices. Each number is
	// either a counter (positive) or a negated child index.
	std::unique_ptr<std::atomic_int32_t[]> m_nodes;
	std::atomic_int32_t m_allocationCounter;
	std::atomic_int32_t m_depth;
	vec3 m_minBound;
	vec3 m_sceneSizeInv;
	float m_sceneScale;
	float m_densityScale;			// 1/#iterations to normalize the counters
	int m_capacity;
	int m_splitCountDensity;	// The number when a node is split must be a
														// multiple of 8 and must grow proportional
														// to #iterations
	bool m_stopFilling;

	// Returns the new value
	int increment_if_positive(int idx) {
		int oldV = m_nodes[idx].load();
		int newV;
		do {
			if(oldV < 0) return oldV;	// Do nothing, the value is a child pointer
			newV = oldV + 1;					// Increment
			// Write if nobody changed the value in between
		} while(!m_nodes[idx].compare_exchange_weak(oldV, newV));
		return newV;
	}

	// Returns the next child pointer or 0
	int split_node_if_necessary(int idx, int count, int currentDepth) {
		// The node must be split if its density gets too high
		if(count >= m_splitCountDensity) {
			// Only one thread is responsible to do the allocation
			if(count == m_splitCountDensity) {
				int child = m_allocationCounter.fetch_add(8);
				if(child >= m_capacity) { // Allocation overflow
					// Avoid overflow of the counter (but keep a large number)
					m_allocationCounter.store(int(m_capacity + 1));
					return 0;
				}
				// We do not know anything about the distribution of photons
				// -> equally distribute. Therefore, all eight children are
				// initilized with SPLIT_FACTOR on set_iteration().
				m_nodes[idx].store(-child);
				// Update depth
				atomic_max(m_depth, currentDepth+1);
				// The current photon is already counted before the split
				// -> return stop
				return 0;
			} else {
				// Spin-lock until the responsible thread has set the child pointer
				int child = m_nodes[idx].load();
				while(child > 0) {
					// Check for allocation overflow
					if(m_allocationCounter.load() > m_capacity)
						return 0;
					child = m_nodes[idx].load();
				}
				return child;
			}
		}
		return count;	// count is already a child
	}
};
\end{lstlisting}

\end{document}